\documentclass[useAMS,usenatbib]{mnras}

\usepackage{graphicx}
\usepackage{amssymb}
\usepackage{amsmath}
\newcommand{\kms}{km~s$^{-1}$}
\newcommand{\jykms}{Jy~km~s$^{-1}$}

\newcommand{\HI}{\ion{H}{i}}
\newcommand{\HII}{\ion{H}{ii}}
\newcommand{\p}{$\pm$}
\newcommand{\sunn}{$_{\odot}$}
\newcounter{qub}

\newcommand{\atoms}{atoms~cm$^{-2}$}

\DeclareRobustCommand{\ion}[2]{%
\relax\ifmmode
\ifx\testbx\f
{\mathrm{#1\,\textsc{#2}}}\else
{\mathrm{#1\,\mathsc{#2}}}\fi
\else\textup{#1\,{\mdseries\textsc{#2}}}%
\fi}

\title[\HI\ study of XMP gas-rich dwarfs in voids -- I]
{\HI\ Studies of Extremely Metal-Poor Dwarfs in Voids -- I}

\author[S. Kurapati, S. A. Pustilnik, E. S. Egorova]
{Sushma Kurapati,$^{1}$\thanks{E-mail: sushma.kurapati@uct.ac.za (SK)}
Simon A. Pustilnik,$^{2}$ Evgeniya S. Egorova$^{3,4}$ \\
\rule{-4pt}{20pt}
$^1$
  Department of Astronomy, University of Cape Town, Rondenbosch 7701, South Africa \\
$^2$ Special Astrophysical Observatory of RAS, Nizhnij Arkhyz,
  Karachai-Circassia 369167, Russia\\
$^3$ Astronomisches Rechen-Institut, Zentrum f\"{u}r Astronomie der Universit\"{a}t Heidelberg, M\"{o}nchhofstra\ss e 12-14, D-69120 Heidelberg, Germany\\
$^4$ Sternberg Astronomical Institute of Moscow State University,
Universitetskij Pr., 13, Moscow, Russia} 

\begin{document}

\label{firstpage}

\date{Accepted  XXX. Received 2023, December 5}

\pagerange{\pageref{firstpage}--\pageref{lastpage}} \pubyear{2023}

\maketitle

\begin{abstract}
We present and discuss the results of the Giant Metrewave Radio Telescope
(GMRT) \HI\ 21-cm line mapping for five {\it isolated} low-mass
(M$_{\rm bary}$$\sim$(2--8)$\times$10$^7$~M\sun) eXtremely Metal Poor (XMP) dwarfs
[12+$\log$(O/H)=7.13-7.28], selected from the Nearby Void Galaxy (NVG) sample.
All the studied void dwarfs show disturbed  morphology in the \HI\ maps with
the angular resolutions of $\sim$11$\arcsec$ to $\sim$40$\arcsec$. We examine the \HI\
morphology and velocity field and the relative orientation of their stellar and gas body spins.
We discuss the overall non-equilibrium state of their gas and the possible origin
and evolution of the studied void dwarfs. The most straightforward interpretation
of the ubiquitous phenomenon of the gas component non-equilibrium state
in these and similar void dwarfs is the cold accretion from the void filaments
and/or minor mergers. The cold gas accretion in voids could be linked to the presence of small filaments that constitute the substructure of voids.
%The important role of the cold accretion in voids can be related to the small filaments,which comprise, as emphasized by Aragon-Calvo and Szalay, the universal void substructure.
\end{abstract}

\begin{keywords}
galaxies: dwarf -- galaxies: evolution -- galaxies: individual:
J0001+3222, J0231+2542, J0306+0028, J1259--1924, J2103--0049 --
galaxies: kinematics and dynamics --  radio lines: galaxies
-- cosmology: large-scale structure of Universe
\end{keywords}

\section[]{INTRODUCTION}
\label{sec:intro}

The study of voids and their galaxies remains one of the current directions
in the understanding of formation and evolution of galaxies and their structures
(see, e.g., review by \citet{vande11, vande16} and references therein). However, the majority of the known studies 
on void galaxies are biased towards the relatively distant voids with typical distances, $D$ $\gtrsim$ 80 -- 200 Mpc. 
The void galaxies they probe are related to the upper part of the galaxy Luminosity Function (LF), namely those with
$M_{\rm B,r} \lesssim -17$ \citep[e.g.,][]{Rojas04,Rojas05,Sorrentino06, wegner19}. For this luminosity range, 
relatively modest differences are found in void galaxies in comparison to their counterparts in denser environments.
%(some cites like Rojas, Hoyle et al., Patiri et al.)

To understand the effect of void environment on a much more numerous population of
the smaller-mass galaxies, one should explore objects in the nearby voids.
The typical low limits on the apparent magnitude of the common redshift surveys are on the
level of $B$,r $\sim$ 18--18.5 mag \citep[e.g.][]{Abazajian09}.
With these limits, in order to probe faint galaxies with luminosities down to M$_{\rm B} \sim$ --12~mag and fainter, one should focus on voids and their galaxies in our nearby surroundings. To this end, \citet{pustilniktepa11} formed a galaxy sample composed of galaxies residing in one of the nearby voids, the Lynx-Cancer void.

The study of this sample, consisting of over a hundred galaxies with a median M$_{\rm B} \sim$ --14, was primarily aimed to determine their evolutionary parameters, specifically the gas metallicity and gas mass fraction (or its substitute, the ratio M(\HI)/L$_{\rm B}$) \citep{Pustilnik11, Perepelitsyna14, Pustilnik16, Pustilnikmartin16}. 
In addition, the \HI\ gas structure and kinematics were studied \citep{ chengalur13, chengalur15, chengalur17, kurapati18, kurapati20, kurapati20a}, along with the photometric parameters of the sample \citep{Perepelitsyna14}. This study covered the great majority of the sample galaxies without any
additional bias. As a result, it was found that void galaxies, as a group, have reduced gas metallicity relative to the reference sample of similar galaxies in the Local Volume \citep{Berg12}, on average by $\sim$40\%, albeit with significant scatter \citep{Pustilnik16, Pustilnik21, Pustilnik24}. Similarly, for the same luminosity, their parameter M(\HI)/L$_{\rm B}$ appeared to be, on average, approximately 40\% higher \citep{Pustilnikmartin16}.

As an additional outcome of this unbiased study, it was found that among the least luminous
void dwarfs there is a substantial fraction of the unusual, eXtremely Metal-Poor
(XMP) galaxies. They have significantly reduced  O/H values for their luminosities 
(by a factor of 2--5), corresponding to 12+$\log$(O/H)=7.0--7.3. Moreover, they have a very high gas
mass fraction (M$_{\rm gas}$/M$_{\rm bary}$ $\sim$ 0.95--0.99).
Furthermore, the majority of this small group of unusual dwarfs display very blue
colours in their outer regions, indicating the ages of the main stellar population in the
range of $\lesssim$(1--4)~Gyr. These properties resemble what one would expect for galaxies in the early phases
of evolution. Interestingly, they also appear to be the best proxies for the nearby so called VYGs
(Very Young Galaxies, suggested in simulations of \citet{Tweed18}).
While such objects are expected to be very rare in the models, they can be
the important probes of the validity of the modern models of galaxy formation.

In order to increase the sample of such unusual dwarfs, many efforts have been made to identify them 
\citep[e.g.,][]{Kniazev03, Izotov06, Izotov07, Izotov18,Izotov19,Sanchez16,Guseva17, Hysu17, Hsyu18, James17, Kojima20}.
However, the overall outcome of these various approaches has been rather limited: only a handful of objects have been identified, and their distances extend up to several hundred Mpc. Several relatively close dwarfs were discovered during extensive searches for the emission-line XMPs in the SDSS database, which are summarized by \citet{Izotov19}. Thanks to their new empirical O/H estimator, best suited for the lowest metallicity \HII\ regions, these authors were able to find the new XMP objects without the temperature-sensitive faint auroral line [O{\sc iii}]$\lambda$4363 in their spectra.

Substantial progress in this direction can be attributed to the formation of a new sample comprising 1,350 galaxies residing in the `Nearby Voids' (D $<$ 25~Mpc) - Nearby Void Galaxies (NVG) sample \citep{pustilnik19}. Utilizing all available information about the NVG objects from public databases and existing literature, \citet{Pustilnik20a} compiled a list of 60 void XMP candidates. These candidates have several properties resembling those of the small group of prototype XMP void dwarfs found in the Lynx-Cancer void. Subsequent spectroscopic observations, conducted using SALT (Southern African Large Telescope)  and BTA (Big Telescope Azimuthal) 
% telescopes 
\citep{Pustilnik20b, Pustilnik21, Pustilnik24}, resulted in the discovery of eleven new void dwarfs with 12+log(O/H) values ranging from 6.98 to 7.21~dex, or gas metallicities of Z(gas) ranging from Z\sun/50 to Z\sun/30.
Additionally, they identified 23 more dwarfs with somewhat higher gas metallicities: Z(gas) $\sim$ Z\sunn/30 -- Z\sunn/20.

Besides, we recently began the complex study of the whole sample of 260 void galaxies
residing within the Local Volume (LV, D $\lesssim$ 11~Mpc) \citep{Pustilnik22b, Pustilnik24}, similar to the unbiased study of the whole sample in the Lynx-Cancer void \citep{Pustilnik16}. This also resulted in several new XMP dwarfs in the nearby voids.

Previous studies of these unusual galaxies have shown that their baryon mass is predominantly composed of atomic gases (\HI\ and He)  \citep[e.g.][]{Pustilnik11, chengalur13, chengalur17, Perepelitsyna14}. To gain a better understanding of their properties, such as dynamical mass, gas morphology, kinematics, and the relationship between gas distribution and star-forming sites, obtaining high-resolution \HI\ 21 cm observations for these galaxies is crucial. Besides, void galaxies in general show signs of filamentary substructures, ongoing gas accretion, as well as interactions with small companions \citep[e.g.][]{Stanionik09, kreckel11, kreckel12, beygu13, Kurapati24}. In our previous GMRT studies of \HI\ gas in several void XMP dwarfs, we have already presented their morphology, kinematics, and related conclusions \citep{Ekta08, Ekta09, ekta10, chengalur13, chengalur15, chengalur17}.  However, the statistics for void XMP dwarfs with high-resolution \HI\ 21 cm observations are still relatively low.

Moreover, extremely gas-rich and metal-poor dwarfs in voids are of interest in the context of studying gas instability and collapse due to their effect on the density threshold for the onset of star formation.  Additionally, interactions and especially merger events involving extremely gas-rich objects are expected to follow different processes compared to galaxies where stars make up the main part of the baryon component. The merging of gas-dominated objects can efficiently dissipate system momentum via shocks. While it undoubtedly occurs in galaxies in the early Universe, this intriguing process remains relatively unexplored, both in observations and models. 

 The main reason for selecting the XMP dwarf sample in this study is an attempt to probe the possible manifestation of cold gas accretion. While this phenomenon is widely discussed and its significant role in galaxy evolution is generally accepted, there is still limited observational support. Gas-rich dwarfs in voids represent one of the best probes to observe the effects of cold accretion. This is primarily due to the much rarer collisions and interactions with galaxy-scale bodies compared to similar objects in walls and filaments. Therefore, in a sufficiently large sample of void dwarfs, one expects that if sizable disturbances exist in the gas distribution and kinematics, they are likely caused not by mutual tidal interactions with other void dwarfs but by cold accretion from small void filaments, as envisioned by \citet{keres05, keres09, aragon13, Dekel13}.The smaller the gravitational potential of a dwarf, the more pronounced the manifestation of the accretion event one can expect. Therefore, we selected a group of galaxies with \mbox{M$_{\rm B}$ $>$--14.2}. In this group, we have a substantial fraction of very rare XMP void dwarfs, which are interesting in the context of their formation, evolution, and star formation.

%Furthermore, 
%voids are promising places to search for tracers of cold gas accretion, %since numerical simulations predict that low density regions at $z \sim 0$ %are dominated by cold gas accretion \citep[e.g.][]{keres05, keres09, %Dekel13}.

In this paper, we present the results of GMRT \HI\ mapping for five {\it isolated} XMP dwarfs located in nearby voids. Three of these dwarfs fall within the Local Volume (LV), allowing for mapping with higher linear resolution. In forthcoming papers, we will extend this study to include ten more gas-rich void XMP dwarfs. 
To investigate star formation in this XMP group, we will analyze high-resolution \HI\ maps and compare them with H$\alpha$ emission distribution, which will be presented later in a separate paper.

The paper is organized as follows: In Section~\ref{sec:sample}, we introduce the sample of observed void XMP dwarfs. In Section~\ref{sec:obs}, we describe the observations and data reduction. In Section~\ref{sec:results}, we present the results of the observations and their analysis. In Section~\ref{sec:dis}, we discuss the results and their implications in a broader context. Finally, in Section~\ref{sec:summary}, we provide a summary and draw our conclusions.

\begin{figure*}
  \centering
\includegraphics[angle=-0,width=3.35cm,clip=]{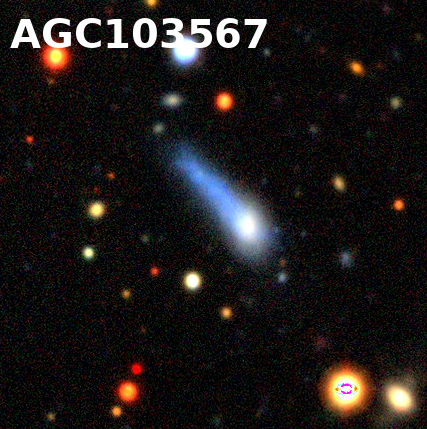}
\includegraphics[angle=-0,width=3.4cm,clip=]{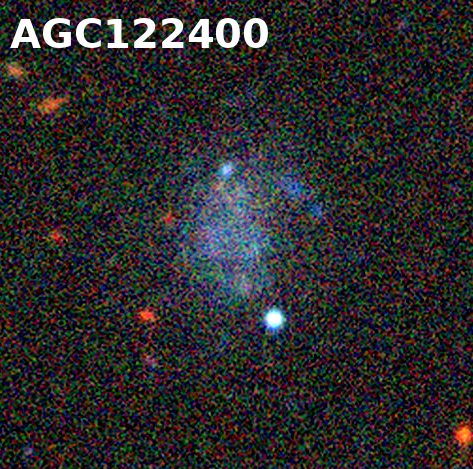}
\includegraphics[angle=-0,width=3.4cm,clip=]{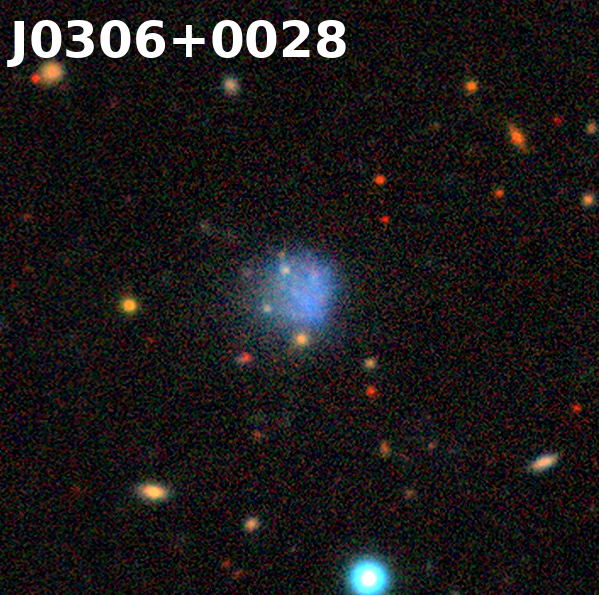}
\includegraphics[angle=-0,width=3.37cm,clip=]{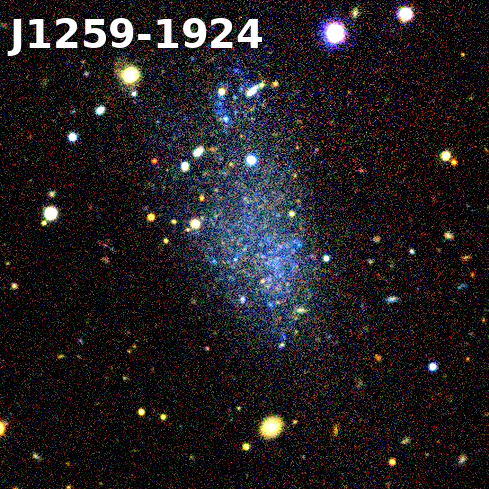}
\includegraphics[angle=-0,width=3.4cm,clip=]{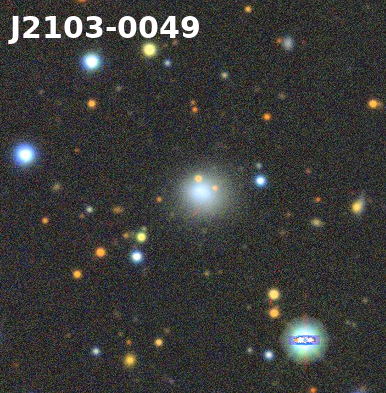}
  \caption{Finding charts for the studied XMP void dwarfs, obtained for all galaxies
   from the Legacy database
   release DR9. The images are 5 kpc on the side. North is up, East
   is to the left. From left to right: J0001+3222, J0231+2542, J0306+0028, J1259-1924 and J2103-0049.
   The strange optical morphology of J0001+3222 is due to the chance projection of the background early-type
   galaxy on to the SW edge of its blue edge-on disc.
}
  \label{fig:FCharts}
 \end{figure*}

\begin{table*}
\caption{Parameters of XMP void dwarfs taken from the literature}
\label{tab:param_literature}
\begin{tabular}{lllllll} \\ \hline
Parameter  & Unit                           & J0001+3222              & J0231+2542       & J0306+0028        & J1259--1924           & J2103--0049            \\         %
			 & 	      & AGC103567               & AGC122400        & PGC1166738        & PGC044681              & PGC1133627             \\ \hline  %
12+log(O/H) & (dex)                     & 7.13\p.06               & 7.20\p.12        & 7.19\p.05         & 7.28\p.05             & 7.20\p.05               \\ % 
$B_{\rm tot}$   &                      & 17.14                   & 18.92            & 17.94             & 17.65                 & 17.44                  \\ 
A$_{\rm B}^{(1)}$  &               & 0.47                    & 0.42             & 0.08              & 0.11                     & 0.11                   \\  
M$_{\rm B}^0$  &               &  --12.86                & --12.45          & --12.67           & --12.01              & --14.07             \\ 
Distance & (Mpc)                        & 9.1$^{(2)}$             & 15.5             & 11.2              & 7.3                  & 17.4                 \\  
V$_{\rm hel}$(\HI)& (\kms)              & 542$\pm$1               & 938$\pm$1        & 710$\pm$1         & 827$\pm$1            & 1411$\pm$1            \\  
S$_{\rm int}$$^{(3)}$  & (\jykms)      & 1.42$\pm$0.04           & 0.89$\pm$0.05    & 0.55$\pm$0.25   & 4.49$\pm$0.14  & 0.95$\pm$0.19          \\ 
W$_\mathrm{50}$$^{(3)}$ & (\kms)        & 32$\pm$2                & 31$\pm$2         & 38$\pm$26       & 38$\pm$2     & 86$\pm$6               \\ 
W$_\mathrm{20}$$^{(3)}$ & (\kms)       & 46$\pm$7                & 69$\pm$13         & 79$\pm$40         & 51$\pm$2       & 104$\pm$9             \\ 
M$_{\rm star}$ & (10$^{7}$ M\sunn)      & 0.155                   & --             & 0.245             & 0.437            & 2.14                  \\  
M(\HI)/L$_{\rm B}$ & (M$_{\odot}$/L$_{\odot}$)  & 1.0             & 3.4              & 1.2               & 7.0        & 1.3                   \\
%
%Opt. size (\arcsec)$^{5}$            & ??$\times$??$^{(2)}$    & ??$\times$?.?    &??$\times$?        &??$\times$??           &??$\times$??            \\         %
%Opt. size (kpc)                      & ?.??$\times$?.??$^{(2)}$& ?.??$\times$?.?? &?.??$\times$?.??   &?.??$\times$?.??       &?.??$\times$?.??        \\         %
%$\mu_{\rm B,c,i}^0$(mag~arcsec$^{-2}$) & 24.:                & 23.2:            & 25.4              & 24.47                 & 24.47                  \\         %
%$f_{\rm gas}$                        & 0.97                    & 0.92:            & 0.98              & 0.??:                 & 0.??                  \\         %
%T(main star population)              & 1--3~Gyr$^{(2)}$        & $\sim$1~Gyr      &                   &                       &                        \\         %
\hline
\multicolumn{7}{p{13.2cm}}{%
%Top part of the Table presents parameters taken from the literature. Bottom part: parameters based on our GMRT and other new data.
(1) -- from NED; (2) -- derived in this paper;
%(3) -- derived from NRT HI profile; % (4) -- corrected for Galactic extinction
%A$_{\rm B}$=0.08; % (5) -- $a \times b$ at $\mu_{\rm B}=$25\fm0~arcsec$^{-2}$;
 (3) --  from published single-dish data  % (8) \citet{DDO68};
% (9) - \citet{Ekta08}; (10) - \citet{IT07};
% (11) - \citet{DDO68_sdss}; (11) - \cite{PaperV}, GMRT flux is 0.725 (see text).
}
\end{tabular}
\end{table*}
\begin{table*}
\caption{Parameters of the GMRT observations}
\label{tab:obspar1}
\begin{tabular}{llllll}
\hline
			 & J0001+3222  & J0231+2542  & J0306+0028 &   J1259-1924 & J2103-0049  \\
			 & AGC103567   & AGC122400   & PGC1166738 &   PGC044681  & PGC1133627  \\
\hline
Date of observations     & 2021 Jun 04 & 2021 Jun 06 & 2021 Jun 05 & 2021 Jul 13 & 2021 Sep 07 \\
			 &             &             & 2021 Jun 07 &             &             \\
Field center R.A.(2000)  &00$^{h}$01$^{m}$06.5$^{s}$  & 02$^{h}$31$^{m}$22.1$^{s}$ &03$^{h}$06$^{m}$46.9$^{s}$  & 12$^{h}$59$^{m}$56.6$^{s}$  & 21$^{h}$03$^{m}$47.2$^{s}$  \\
Field center Dec.(2000)  &32$^{o}$22$^{'}$41.0$^{"}$  & 25$^{o}$42$^{'}$45.0$^{"}$ &00$^{o}$28$^{'}$11.0$^{"}$  &--19$^{o}$24$^{'}$41.0$^{"}$ &--00$^{o}$49$^{'}$50.0$^{"}$  \\
Central Velocity (\kms)  & 542  & 938  & 700 & 827 & 1411 \\
Time on-source  (h)      &$\sim$9  & $\sim$9 & $\sim$7+7 & $\sim$9 & $\sim$9 \\
Number of channels       & 512 & 512  & 512 & 512 & 512 \\
Channel separation (\kms)& $\sim$1.7 & $\sim$1.7 & $\sim$1.7 & $\sim$1.7 & $\sim$1.7 \\
Flux Calibrators         & 3C48      & 3C48      & 3C48     & 3C147 & 3C48 \\
Phase Calibrators        & 0029+349  & 0237+288  & 0323+055 & 1351-148  & 2136+006 \\
Resolution (arcsec$^{2}$) (rms (mJy~Bm$^{-1}$)) & 44.8~$\times$~32.5 (1.6) & 43.5~$\times$~34 (1.6) & 43.5~$\times$~36.5 (1.8)  & 46.5~$\times$~33.5 (1.8) & 45.5~$\times$~37.5 (1.6) \\
						& 17.1~$\times$~12.9 (1.1) & 16.5~$\times$~13 (1.1) & 17~$\times$~13.5 (1.4)  & 16~$\times$~12 (1.4) & 20~$\times$~15 (1.1) \\
						& 12.9~$\times$~9.1 (1.0) & 12~$\times$~9.5 (1.0) & 12.5~$\times$~10 (1.2)  & 11~$\times$~8.5 (1.2) & 14.5~$\times$~9.5 (1.0) \\
%                                               & 8~$\times$~7   (0.9) & 8~$\times$~7   (0.9) & 7~$\times$~5          & 7~$\times$~5         & 8~$\times$~7   (0.9) \\
\hline
\hline
\end{tabular}
\end {table*}

\section[]{SAMPLE}
\label{sec:sample}

The entire sample of XMP dwarfs in the Nearby Voids, for which we obtained \HI\ 21 cm observations, includes approximately 15 gas-rich galaxies observed at GMRT. These XMP dwarfs were identified based on the results of studies conducted in both the Lynx-Cancer void \citep{Pustilnik16} and the `Nearby Voids' sample \citep[][]{Pustilnik20b, Pustilnik21, Pustilnik24}. In this paper, we present the results for 5 dwarfs observed at GMRT from June to September 2021. The remaining XMP and/or gas-rich void dwarfs, observed in other periods, will be presented in subsequent papers.

The summary of their main properties, known from the literature and from our previous results, is presented in the Table~\ref{tab:param_literature}. The rows are as follows:\\
Row 1 - gas `metallicity' 12+log(O/H) (in dex). This is typically determined using the electron temperature method ($T_{\rm e}$ method) if the principal faint auroral line [O{\sc iii}] $\lambda$4363~\AA\ can be reliably measured. In cases where this line is too faint or undetectable, as for all dwarfs in this work, various empirical O/H estimators are employed.
We used a robust empirical O/H estimator, based on the intensities of strong oxygen lines \citep{Izotov19}, specifically suited
for the lowest metallicity range, 12+log(O/H) $\lesssim$7.5, which has an accuracy of 0.04~dex \citep{Izotov19, Pustilnik21}. All the values of 12+log(O/H) shown in this table are from \citet{Pustilnik20b, Pustilnik21, Pustilnik24}. For galaxy J1259--1924, which had relatively low signal-to-noise spectra, the adopted value of 12+log(O/H) is the average of two different \HII-regions on spectra from \citet{Pustilnik20b, Pustilnik24}.\\
Row 2 - the total B magnitude, not corrected for A$_{B}$, obtained by transformation from the total g and r, according to the formulae given in Lupton et al (2005)\footnote{http://www.sdss.org/dr5/algorithms\\/sdssUBVRITransform.html\#Lupton2005};\\
Row 3 - A$_{\rm B}$, the Galactic extinction in B band; \\
Row 4 - the absolute blue magnitude corrected for the Galactic extinction; \\
Row 5 - the adopted distance in Mpc. The distance for galaxy J1259--1924 is determined via the Tip of RGB method based on the photometry of individual stars on the HST (Hubble Space Telescope) images \citep{Karachentsev17}. For the remaining four dwarfs, the distance was estimated from their radial velocities according to the kinematic model by \citet{Tully08}. This model accounts for the motions of the Local Sheet towards the nearby attractor in the direction of the Virgo cluster and away from the center of the Local Void \citep[see][for more details]{pustilnik19}; \\
Row 6 - the heliocentric velocity, obtained from \HI\ profile in this paper; \\
Row 7 - Integrated flux, $S_{\rm int}$ measured from single dish \HI\ observations ;\\
Row 8 and 9 - Line widths W$_{50}$ and W$_{20}$ in \kms measured from single dish \HI \ observations . These values represent the measured linewidths at 50 \% and
20 \% of the peak flux, respectively; \\
Row 10 - stellar mass in M$_{\odot}$; \\
Row 11 - ratio of \HI\ mass to blue luminosity, M(\HI)/L$_{B}$ in solar units. 

In Figure~\ref{fig:FCharts}, we present the deepest available color images of the five
studied void XMP dwarfs. All the images were obtained from DR9 or DR10 of the DESI Legacy Imaging Surveys \citep{Dey19}, except for J2103--0049, for which we selected the deep images from SDSS Stripe82 \citep{Fliri16}. This figure clearly shows the optical sizes and morphologies of the XMP dwarfs, which may not be as distinct in the subsequent figures where \HI\ maps are overlaid.

\begin{table*}
\caption{Properties of XMP dwarfs derived using GMRT \HI\ 21 cm observations}
\label{tab:param_GMRT}
\begin{tabular}{lllllll} \\ \hline
Parameter  & Unit  & J0001+3222  & J0231+2542       & J0306+0028        & J1259--1924       & J2103--0049     \\       
	& 	      & AGC103567   & AGC122400        & PGC1166738        & PGC044681         & PGC1133627       \\ \hline  %
\hline
S$_{\rm int}$    &  (\jykms)   & 0.98$\pm$0.11    & 0.57$\pm$0.09    & 0.35$\pm$0.04     & 4.28$\pm$0.45  & 0.84$\pm$0.10     \\         %
W$_\mathrm{50}$  &  (\kms)     & 25$\pm$4               & 26$\pm$2        & 13$\pm$4        & 39$\pm$4           & 86$\pm$12   \\         
W$_\mathrm{20}$  & (\kms)      & 35$\pm$6               & 31$\pm$3       & 24$\pm$6        & 50$\pm$6           & 98$\pm$19    \\ 
b/a               &            & 0.59             & 0.72      & 0.54      & 0.55        & 0.32    \\
$i_{\rm HI}$     & ($^{\circ}$)& 68               & 49        & 67        & 64          & 84      \\
$\sigma_{\rm corr}$&  (\kms)    & 6--7             & 8--9    & 6--8       & 6--10      & 11--12  \\ 
%V$_\mathrm{rot}$~$sin~{i}$ & (\kms) & 34          & 22        & 23       & 25         & 60      \\   
M(\HI)          & (10$^{7}$ M\sunn) & 2.8         & 4.1       & 1.3       & 5.3         & 9.3     \\ 
%M(\HI)        & (10$^{7}$ M\sunn) & 2.8          & 5.0       & 1.9       & 6.3         & 13.4             \\     
M$_{\rm gas}$ & (10$^{7} M_{\odot}$) & 3.6        & 5.3       & 1.7       & 6.9         & 12.1     \\ 
R$_{\HI, m}^{a}$    &  (kpc)      & 1.2             & 1.7      &   1.1    & 2.2         & 7.3   \\
R$_{\HI, d}^{b}$    &  (kpc)      & 1.3             & 1.7      &   0.8    & 2.1         & --   \\
M$_{\rm dyn}^{c,e}$ & (10$^{8} M_{\odot}$) & 1.5      & 4.2     & 2.1      & 5.9         & 77.4 \\    
M$_{\rm dyn}^{d,e}$ & (10$^{8} M_{\odot}$) & 1.6      & 9.3      & 1.6      & 5.6         & --     \\    
M(\HI)/L$_{\rm B}$  & (M\sunn/L\sunn)  & 1.0   & 2.8      & 0.9   & 6.6   & 1.7    \\ %
% M(\HI)/L$_{\rm B}$ & (M\sunn/L\sunn)   & 1.3  & 2.3   & 1.1    & 3.1  & 2.6    \\ %

%$f_{\rm gas}$                     & 0.97          & 0.92:            & 0.98     & 0.??:   & 0.??                  \\         %
%T(main star population)         & 1--3~Gyr$^{(2)}$   & $\sim$1~Gyr    &        &            &                        \\      %
\hline
\multicolumn{7}{p{13.2cm}}{ $^{a}$ Radius of the galaxy based on moment maps; $^{b}$ Radius of the galaxy where the inclination corrected surface density is 1 M$_{\odot}$ pc$^{-2}$; $^{c}$ Dynamic mass of the galaxy based on R$_{\HI, m}$;  $^{d}$ Dynamic mass of the galaxy based on R$_{\HI, d}$; $^{e}$ The dynamic mass estimate (using equation \ref{eq:2}) assumes that the velocity amplitude is due to the regular rotation. %Since for all our dwarfs we find various indications of  the `unsettled' gas, the derived values of dynamical mass are somewhat conditional and bear the additional uncertainties. For the outlying value of M$_{\rm dyn}$ in J2103--0049, see Sections~\ref{ssec:J2103}, \ref{ssec:global_parameters} for alternative interpretations. 
}
\end{tabular}
\end{table*}

\begin{figure*}
\centering
\includegraphics[width=0.32\linewidth]{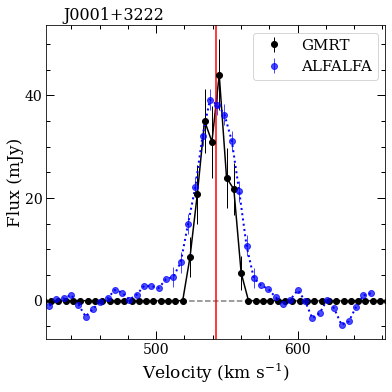}
\includegraphics[width=0.32\linewidth]{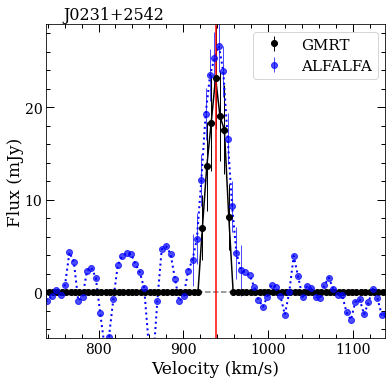}
\includegraphics[width=0.32\linewidth]{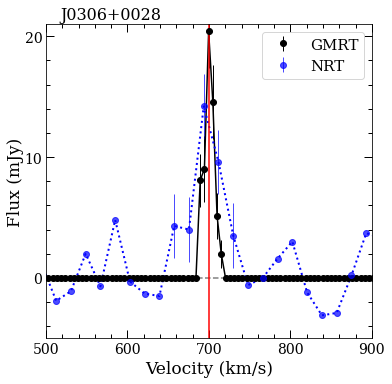}
\includegraphics[width=0.32\linewidth]{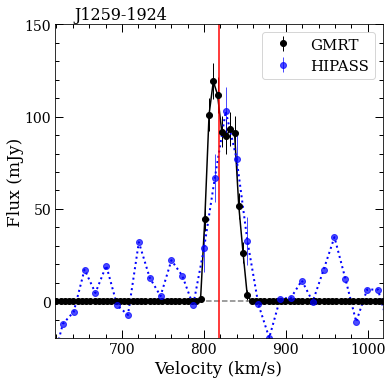}
\includegraphics[width=0.32\linewidth]{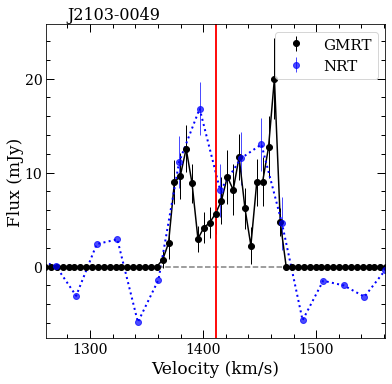}
\caption{\HI\ spectra of J0001+3222, J0231+2542, J0306+0028, J1259-1924, and J2103-0049 respectively. Black solid line indicates the spectra from GMRT observations from this work. Blue dotted line shows the \HI\ spectra from single dish observations from ALFALFA survey \citep{Haynes18}, NRT survey \citep{vandriel16}, and HIPASS survey \citep{Meyer04}.}
\label{fig:HI_prof}
\end{figure*}

\section{OBSERVATIONS AND DATA ANALYSIS}
%\section[]{OBSERVATIONS AND DATA REDUCTION}
\label{sec:obs}

The GMRT \HI\ 21-cm observations of the 5 void XMP galaxies presented in this paper were conducted between June and September 2021. 
 We used the GMRT Software Back-end (GSB) setup, which is centered at the heliocentric redshift of each galaxy, providing a total bandwidth of 4.17 MHz ($\sim$ 890 \kms) divided into 512 channels, resulting in a velocity resolution of 1.74 \kms. All sources were observed for 9 hours, except for the faintest galaxy (J0306+0028), which was observed for $\sim$ 14 hours. The parameters of the GMRT observations are summarized in Table~\ref{tab:obspar1}.

Initial flagging and calibration were performed using the stand-alone pipeline `{\texttt{flagcal}}' \citep{Prasad12}, and subsequent processing was carried out using standard tasks within the {\texttt{AIPS}} package. The `CVEL' task was applied to the calibrated data sets to correct for the Doppler shift due to Earth's motion. The galaxy J0306+0028 was observed over two sessions, which were combined using the task `DBCON' before imaging. A continuum image was generated using the line-free channels, and the corresponding clean components were subtracted using the `UVSUB' task. Any residual continuum was subtracted in the image plane using `IMLIN'. Data cubes were generated by averaging three channels, resulting in a channel width of approximately $\sim$5.2 \kms. These cubes, produced at multiple resolutions, leverage the hybrid configuration of GMRT to create both high and low-angular resolution images from a single observation. These cubes were produced for various (u,v) ranges, including 0–5 k$\lambda$, 0–15 k$\lambda$, and 0–20 k$\lambda$, corresponding to a beam FWHM of roughly 40$\arcsec$, 15$\arcsec$, and 10$\arcsec$ respectively.

We use the Source Finding Application (SoFiA2; \url{https://github.com/SoFiA-Admin/SoFiA-2}) pipeline  \citep{serra15,Westmeier21}, which is designed for automatic H{\sc i} source finding  in interferometric spectral line data to detect the \HI\ emission in data cubes and to create moment maps. We use the `smooth+clip' algorithm which iteratively smooths the data cube on multiple spatial and spectral scales to obtain significant emission above a user-specified detection threshold on each scale. Applying a 4 $\sigma$ clip level, we extract \HI\ parameters of each galaxy including moment maps. We did not detect any companion galaxies associated with any of the galaxies.

\section{Results}
\label{sec:results}
\subsection[]{\HI\ integrated parameters}
\label{sec:HI_parameters}

%In this section, we compare the global \HI profiles and their derived parameters, including the integrated flux $F$(\HI), central velocity, and the widths at levels of 0.5 and 0.2 of the peak flux density (W${\mathrm 50}$ and W${\mathrm 20}$), obtained from GMRT observations in this work with data from earlier single dish observations.

In this section, we compare the global \HI\ profiles and their derived parameters obtained from GMRT interferometric observations in this study, with those from previous single-dish observations. The \HI\ spectra were extracted using the SoFiA 3D mask in regions where the \HI\ emission is detected, and integrated fluxes are measured from these spectra.  The error on the flux density is dominated by the calibration uncertainties, which for the GMRT are typically $\sim$ 10 percent.

%The \HI\ spectra were extracted from the unsmoothed cubes in the regions where the \HI\ emission is detected. Integrated fluxes are measured from these integrated global profiles. 
We present the global \HI\ emission of all the galaxies in Fig. \ref{fig:HI_prof}. The black solid lines indicate the \HI\ profiles obtained from the GMRT observations, while the blue dotted lines show the integrated single-dish profiles \citep{Haynes18, vandriel16, Huchtmeier00}. The single-dish \HI\ observations of J0001+3222 and J0231+2542 are from the blind \HI\ ALFALFA survey \citep{Haynes18}, observed by the Arecibo telescope, which has a resolution of approximately 3.5 arcmin. The single-dish \HI\ profiles of J0306+0028 (cataloged as NIBLES0347) and J2103--0049 (cataloged as NIBLES2400) are taken from \citet{vandriel16}, observed by the Nançay Radio Telescope (NRT). The \HI\ profile of galaxy J1259--1924 was obtained from the HIPASS survey \citep{Meyer04}, observed with the Parkes telescope.

 For the GMRT data, the error bars on the flux in each channel were calculated using the formula:
\begin{equation}
    \sigma_{\text{channel}} = \frac{\sigma_{\text{rms}}}{\sqrt{n_{\text{px}}}} \sqrt{N_{\text{px\_em}}}
\end{equation}

where $\sigma_{\text{rms}}$ is the  RMS noise in mJy beam$^{-1}$ per channel,  \(n_{\text{px}}\) is the number of pixels per beam, and \(N_{\text{px\_em}}\) is the number of pixels with \HI\ emission in that channel. For the single dish spectrum, the error bars were determined by identifying non-emission channels and calculating the standard deviation of the flux values in these channels. Please note that the error bars in Figure 2 do not include the calibration uncertainties.

The GMRT profiles for all the galaxies, except J2103--0049, match well with their corresponding single-dish \HI\ profiles within the error bars. The integrated flux $S_{\rm int}$ measurements coincide within the error bars, although the single-dish fluxes are slightly higher than the GMRT fluxes, as expected. However, for J0001+3222 and J0231+2542, the ALFALFA flux is approximately 45--55$\%$ higher than the GMRT flux. The ALFALFA blind survey has an effective beam width of approximately 3.5 arcmin. This difference may be attributed to missing flux, i.e., a large fraction of \HI\ being in an extended distribution that is resolved out. Another possibility is the presence of multiple sources within the Arecibo beam. However, the latter scenario seems less likely, as an \HI\ source with a flux comparable to that of J0001+3222 or J0231+2542 should have been detected by GMRT, considering the primary beam of GMRT, which has a FWHM of $\sim$ 22 arcmin. For the galaxy J2103--0049, although the integrated flux matches within the error bars, the spectrum looks different. 

The line widths ($W_{50}$ and $W_{20}$) in \kms\ measured for GMRT profiles are taken from the SoFiA output. The error bars on these line widths ($W_{50}$ and $W_{20}$) are calculated using the method described by \citet{Schneider86}:
\begin{equation}
    \delta(W_{50}) = 3.0 \times \frac{W_{20} - W_{50}}{S/N}
\end{equation}
\begin{equation}
    \delta(W_{20}) = 4.7 \times \frac{W_{20} - W_{50}}{S/N}
\end{equation}
where \(S/N\) is the signal-to-noise ratio. The line width measurements with GMRT observations match with single dish observations within the error bars for all the galaxies except for J0306+0028, where NRT observations showing higher line widths. However, it is important to note that the error bars on NRT line widths are high, and the single-dish spectrum also appears noisier.

  In Table~\ref{tab:param_GMRT}, we present all the \HI\ integrated parameters obtained from GMRT observations in this work.  The rows are as follows:\\
Row 1 - Integrated flux, $S_{\rm int}$ measured from the integrated global profiles and the error on the integrated flux $S$ was obtained by taking the quadrature of RMS noise and uncertainty due to absolute flux calibration; \\
Row 2 and 3 -  Line widths W$_{50}$ and W$_{20}$ in \kms measured from GMRT profiles\\

Row 4 and 5 - b/a, minor axis to major axis ratio and \HI\ morphological inclination; The values of minor axis and major axis were obtained from the SoFiA output catalogues, which were derived by fitting ellipses to isophotes of \HI\ maps. These ratios and inclinations were obtained at an angular resolution of 15\arcsec. The inclinations (\(i_{\rm HI}\)) were estimated based on the ratio of minor to major axis, including a correction factor dependent on the intrinsic axial ratio distribution from a sample of galaxies, as described in \citet{Staveley-Smith92}: 
\begin{equation} \label{eq:1}
    \cos^{2}(i) = \frac{\left(\frac{b}{a}\right)^{2} - q_{\mathrm 0}^{2}}{1 - q^{2}_{\mathrm 0}}
\end{equation}
where \(q_{\mathrm 0}\) is the intrinsic axis ratio. We assume an intrinsic thickness of \(q_{\mathrm 0} \sim 0.4\) \citep{Roychowdhury13, Sanchez10} for these faint dwarfs. \\
Row 6 - The velocity dispersion, after correcting for the contribution of the velocity gradient convolved with the beam, is in units of \kms; \\
%Row 6 - Projected rotation velocity in \kms \\
Row 7 and 8 -  measured \HI\ mass and gas mass (corrected for helium) in units of 10$^{7}$~M\sunn.  We estimate the \HI\ mass using the standard formula  $M_{\rm HI} = 2.356\times 10^{5} D^{2} S_{\rm int} \ {\rm M}_{\rm \odot}$, where $D$ is the distance in Mpc, $S_{\rm int}$ is the integrated flux density in 
Jy~\kms\ \citep{Roberts62}. We adopt distances and integrated flux of \HI\ presented  in Table~\ref{tab:param_literature}\\
Row 9 -  The radius of the galaxy (R$_{\HI, m}$) is based on the extent of the moment maps in kpc. To account for the beam contribution to the size of the galaxy, we have subtracted the beam size from the measured \HI\ diameter in quadrature. These values are derived homogeneously for all studied galaxies from the moment maps with FWHM $\sim$15$\arcsec$. For two galaxies that are well resolved with a beam size of FWHM $\sim$40$\arcsec$, we provide alternative estimates in Sections~\ref{ssec:J1259} and \ref{ssec:J2103} to better understand how they can vary. \\
Row 10 -  Radius of the galaxy R$_{\HI, d}$ in kpc is  determined by fitting ellipses to the contour of the inclination-corrected \HI\ column density map  ($\sim$15$\arcsec$ resolution) corresponding to 1 M$_{\odot}$ pc$^{-2}$.  \\
Row 11 and 12 - Estimated dynamical mass using the radius as derived in Row 8 and Row 9 respectively. We use a simple dynamical mass modeling approach since these galaxies have disturbed velocity fields. To account for the effect of velocity dispersion on the line width, we correct the observed line width using the method described by \citet{Tully85}. The corrected line width is calculated as:
\[
W_{\rm rot}^2 = W_{20}^2 + W_t^2 - 2W_{20}W_t \left[1 - \exp\left(-\left(\frac{W_{20}}{W_c}\right)^2\right)\right]
\]
\[
- 2W_t^2 \exp\left(-\left(\frac{W_{20}}{W_c}\right)^2\right)
\]

In this equation, $W_t$ represents the velocity dispersion component (random component), and $W_c$ is the transition parameter, which is 120 \kms \citep{Tully85}. $W_{20}$ (in \kms) is the line width measured from GMRT profiles.
The dynamical mass is then estimated (following \citealt{deblok14, Guo20}) using:
\begin{equation} \label{eq:2}
    M_{\rm dyn} [M_{\odot}] = 2.31 \times 10^{5} \left(\frac{W_{\rm rot}}{2 \sin(i)}\right)^{2} R_{\rm HI}
\end{equation}
In this equation, $W_{\rm rot}$ is the corrected line width, $i$ is the inclination angle, and $R_{\rm HI}$ is the radius (in kpc) as determined in the previous rows.\\
%We estimate the dynamical mass (following \citet{deblok14, Guo20}) using:
%\begin{equation} \label{eq:2}
%    M_{\rm dyn} [M_{\odot}] = 2.31 \times 10^{5} (W_{20}/2sin(i))^{2} R_{\rm HI}
%\end{equation}
%where $W_{\rm 20}$ (in \kms) is the line width measured from GMRT profiles, $i$ is the inclination angle, and R$_{\rm HI}$ is the radius(in kpc) as determined in the previous rows.
Row 13 - the ratio of \HI\ mass to blue luminosity, M(\HI)/L$_{B}$ in solar units. 

\begin{figure*}
\centering
 \includegraphics[width=0.95\linewidth]{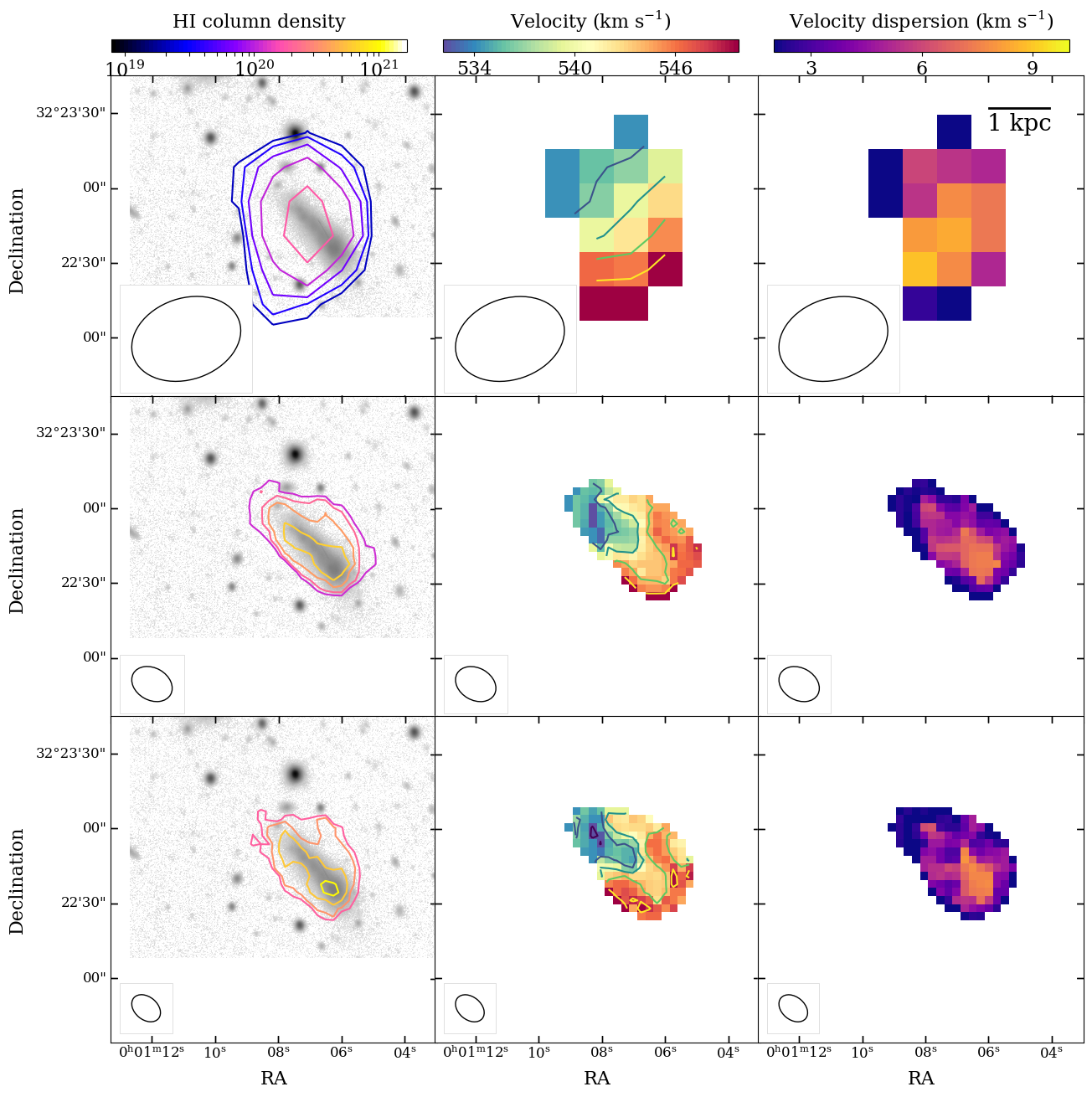}
  \caption{\label{fig:HImaps1}  The \HI\ moment maps for the XMP void dwarf J0001+3222 at three distinct resolutions: 40'' (top row), 15'' (middle row), and 11'' (bottom row). Left panels show the \HI\ contours overlaid on DECaLS g-band optical images. The plotted \HI\ contours are 2$\times$10$^{19}$  $\times$  ($\sqrt{3})^{n}$~\atoms\ , n=0, 1, 2, 3,... at 40$\arcsec$, 1.1$\times$10$^{20}$  $\times$  ($\sqrt{3})^{n}$~\atoms\ , n=0, 1, 2, 3,... at 15$\arcsec$, and 2.0$\times$10$^{20}$  $\times$  ($\sqrt{3})^{n}$~\atoms\ , n=0, 1, 2, 3,... at 11$\arcsec$. Middle panels display velocity fields (isovelocity contours are spaced at intervals of 4 \kms), while right panels present  Moment 2 maps. The Gaussian beam is shown on the bottom left of each map.}
\end{figure*}

\begin{figure*}
 \centering
 \includegraphics[width=0.95\linewidth]{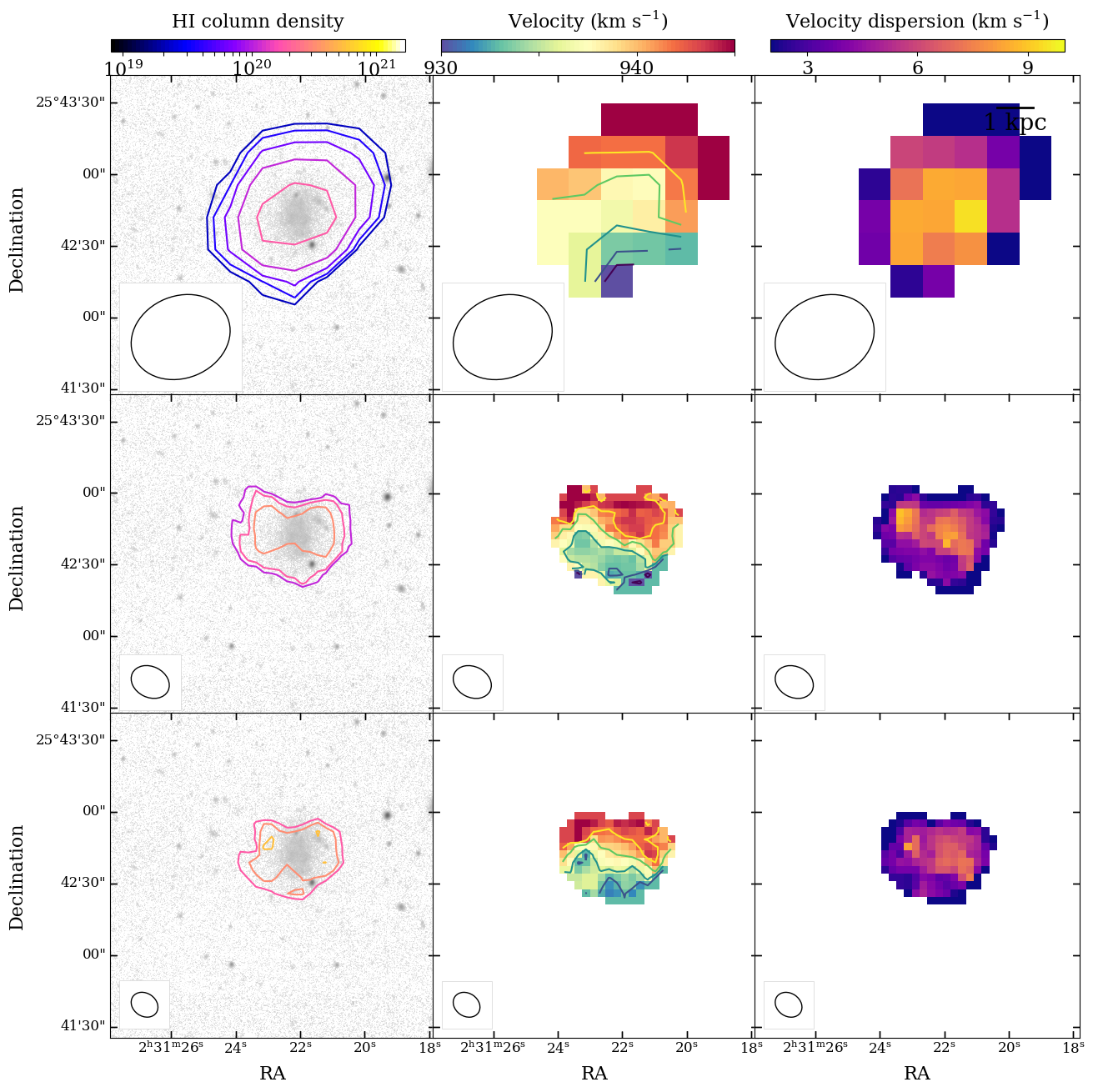}
  \caption{\label{fig:HImaps2} The \HI\ moment maps for the XMP void dwarf J0231+2542. Panels and symbols as in Figure \ref{fig:HImaps1}. The \HI\ contours overlaid on Legacy g-band image in left panels begin at 3$\sigma$, i.e. 2.1 $\times$10$^{19}$, 1.0 $\times$10$^{20}$, and 1.8 $\times$10$^{20}$~\atoms\ at  40$\arcsec$', 15$\arcsec$', and 11$\arcsec$ resolution respectively, and increase by factor of $\sqrt{3}$. Isovelocity contours in the middle panels are spaced at 3~\kms. }
\end{figure*}

\subsection{\HI\ morphology and kinematics}
\label{sec:morpho}

\begin{figure*}
 \centering
  \includegraphics[width=0.95\linewidth]{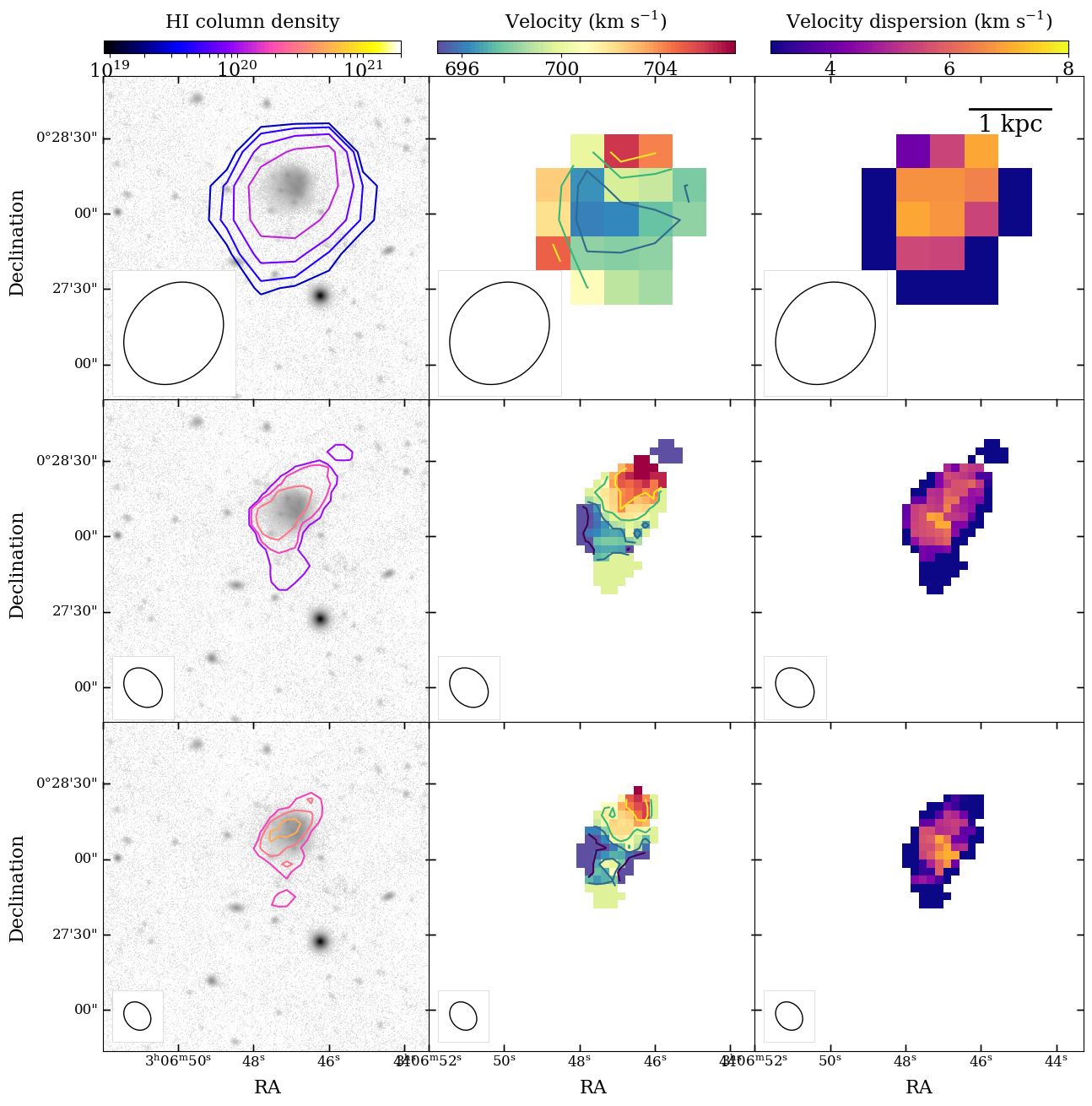}
  \caption{\label{fig:HImaps3} The \HI\ moment maps for XMP void dwarf J0306+0028. Panels and symbols as in Figure \ref{fig:HImaps1}. The \HI\ contours overlaid on Legacy g-band image in left panels begin at 3$\sigma$, i.e. 2.4 $\times$10$^{19}$, 1.0 $\times$10$^{20}$, and 1.7 $\times$10$^{20}$~\atoms\ at  40$\arcsec$', 15$\arcsec$', and 11$\arcsec$ resolution respectively, and increase by factor of $\sqrt{3}$. Isovelocity contours in the middle panels are spaced at 3~\kms. }
\end{figure*}

One of the most prominent features in the morphology of the majority of void XMP gas-rich dwarfs, mapped in the \HI\ line so far \citep[e.g.][]{Ekta08, Ekta09, chengalur15, chengalur17, kurapati18}, as well as seen in this paper (see Figures \ref{fig:HImaps1}-\ref{fig:HImaps5}), is the presence of clear disturbances in their \HI\ density distributions and velocity fields. When these void dwarfs enter to pairs or triplets, such a disturbance looks to be expectable. However, in a number of published apparently isolated void XMP dwarfs such the unsettled gas state requires the study of the reasons of this phenomenon. These disturbances can sometimes be smoothed out in the lowest resolution maps. However, higher-resolution maps reveal many disturbed features. In the following subsections, we provide a more detailed description of the GMRT \HI\ maps obtained for each of the five studied XMP void dwarfs.

\subsubsection{J0001+3222}
\label{ssec:J0001}

%The first information on the integrated properties of  \HI\ emission in this LV XMP dwarf was presented in the blind \HI\ survey ALFALFA \citep{Haynes18} at Arecibo telescope with the beamwidth of FWHM $\sim$3.5~arcmin. We compare them with the GMRT data in Section~\ref{sec:dis}.% See the respective profile in Figure~\ref{fig:HI_prof}.

In the top row of Figure~\ref{fig:HImaps1} (left-hand panel), we show the \HI\ distribution ($\sim$ 40$\arcsec$ ) of the galaxy J0001+3222 overlaid on its optical ($g$-band) image taken from the Legacy surveys database \citep{Dey19}. As explained in \citet{Pustilnik21}, its tadpole appearance is due to
the chance projection of a distant red galaxy on to the edge of the nearby void blue almost edge-on XMP disc. There are no apparent galaxies in the neighborhood within approximately $\sim$60~kpc in projection ($\sim$20$\arcmin$), which could significantly disturb J0001+3222. Therefore, as a first approximation, this galaxy can be considered relatively isolated.

To assess the local environment, we initially examined galaxies with known redshifts within a radius of 30$\arcmin$ using the HyperLEDA database, but did not find any. Subsequently, we carefully examined its environments for the potential fainter counterparts using the
deepest colour images available in the Legacy survey. Four possible blue or bluish objects, ranked by their angular distance to J0001+3222, are situated at 3.3, 6.4, 6.4, and 15.3$\arcmin$, or approximately 10, 19, 19, and 45~kpc, respectively. Their B-band magnitudes, estimated from the SDSS DR16 catalog g and r magnitudes, are as follows: 20.53, 18.06, 19.59, and 17.75.

Assuming a gas-mass to blue luminosity ratio of \mbox{M(\HI)/L$_{\rm B}$ $\sim$ 1,} a typical value for late-type dwarfs in a nearby void and in the denser environments in the Local Volume \citep[e.g.][]{Pustilnikmartin16}, and accounting for the Milky Way extinction A$_{\rm B} \sim$ 0.2 mag, one can estimate the expected \HI\ flux at their positions and the likelihood of detecting them in our GMRT observations. The expected values of $S_{\rm int}$ for these galaxies are 0.05, 0.42, 0.11, and 0.6~\jykms, respectively. In principle, only the two brighter galaxies could be detected, given that the main target with $S_{\rm int}$ $\sim$1.0~\jykms\ is detected with a signal-to-noise ratio of $\sim$20. However, when considering the substantial offset of these objects from the primary beam center and the resulting loss of their flux, one would expect their detection to be at a level of $\lesssim$ 0.2~\jykms\ if they have radial velocities within the range covered by the observations, that is 542$\pm$380~\kms. 
%and are thus situated in the same cosmic `cell'. 
Due to their low expected \HI\ integrated flux, one can not exclude their proximity to the target void XMP dwarf. Nevertheless, the combination of their expected masses and mutual distances suggests that their tidal effect on the target galaxy would be relatively small. 
%({\color{red} NEED an example with the estimate of the strength of the tidal action}).
% position of J000106.5+322241 is RA=0.27593, DEC=32.53763
% In fact there is a similary-looking galaxy (elong.disc)  at ~15.3 arcmin to NE (~45 kpc) with
% 1) RA=0.4758, DEC=32.5384 = SDSS J000154.20+323217.8 = GALEXASC J000154.06+323218.7
% (u=18.70+-0.05; g=17.46+-0.01; r=17.27+-0.01; i=17.28+-0.01' m_NUV=19.41, m_FUV=19.73; translate to B~17.75
% there are 3 more bluish amorph.galaxies:
% 2) at RA,Dec 0.21021, 32.4666 at ~6.4' (~19 kpc), CModel mags: u=19.01, g=18.02, r=17.71, i=17.43 => B=18.06
% 3) at RA,Dec 0.21165, 32.39094 at ~3.3' (~10 kpc) g=20.19 r=19.87 => B~20.53
% 4) at RA,Dec 0.15772, 32.41664 at ~6.4' (~19 kpc) g=19.24 r=18.84 => B~19.59

%{Despite the available data do not provide evidence for the presence of a substantial disturber, the \HI\ morphology of J0001+3222, \textcolor{red}{as seen in the Moment 0 (MOM0) map with a beam of $\sim$40$\arcsec$, is rather asymmetric.  }With this beam size the whole \HI\ body of the galaxy is barely resolved. \textcolor{red}{ However, its major axis is clearly inclined relative to the stellar body. A substantial appendage on the NE side is also evident.} The extent of the \HI\ body is only about twice as large at the column densities of $\sim$10$^{19}$~\atoms, compared to that of the optical disc, as seen in the deepest images.}

Despite the available data not providing evidence for a substantial disturber, the \HI\ morphology of J0001+3222 is rather asymmetric. With a beam size of $\sim$40$\arcsec$, the \HI\ body of the galaxy is barely resolved. The extent of the \HI\ body is only about twice as large at column densities of $\sim$10$^{19}$~\atoms, compared to that of the optical disc, as seen in the deepest images.

In the maps with the higher angular resolutions (the second and the third rows of Fig. \ref{fig:HImaps1}) with the effective FWHM beamwidths of $\sim$15$\arcsec$ (17$\arcsec$ $\times$13$\arcsec$) and $\sim$11$\arcsec$ (12.5$\arcsec$ $\times$9$\arcsec$), respectively, the disturbed morphology and velocity field are even more distinctly visible. Namely, there are two peaks of density along the main body (both within the optical `disc'). The related maps of gas radial velocities also show clear irregularities indicative of non-equilibrium state.  However, these disturbed structures little affect the velocity dispersion, as one can see in MOM2 maps. The observed velocity dispersion ($\sigma_{\rm obs}$) of  6--9~\kms\ is rather typical of dwarf irregular galaxies.  After correction for the effective channel width of 5.1~\kms\ and the contribution of the velocity gradient convolved with the beam, namely the term of 0.5$\times$FWHM$\times$$dV/dr$ \citep[following][]{Begum04}, here of the order of 3--4~\kms,  $\sigma_{\rm true}$ does not exceed 6--7~\kms.

 %Despite the significantly disturbed appearance of the velocity field, a relatively regular velocity gradient is observed in the two higher resolution MOM1 maps, approximately from the NE to the SW. The full amplitude of the velocity change does not exceed $\sim$20~\kms. This is in contrast with the broader velocity range visible in the integrated \HI\ profiles, especially for the ALFALFA one, as presented in Figure~\ref{fig:HI_prof}, where a velocity range of $\sim$32~\kms\ (W$_\mathrm {50}$) and $\sim$46~\kms\ (W$_\mathrm {20}$) is observed. By  adopting the latter value and the estimated inclination angle $i$ = 68\degr\ (using eqn \ref{eq:1}), we derive V$_{\rm rot}$ to be 23~\kms. The full extent of \HI\ gas, as seen on the MOM0 maps of Figure~\ref{fig:HImaps1}, is $\sim$60$\arcsec$ (or 2.65~kpc).

Despite the significantly disturbed appearance of the velocity field, a relatively regular velocity gradient is observed in the two higher resolution MOM1 maps, approximately from the NE to the SW. The full amplitude of the velocity change does not exceed $\sim$20~\kms. Using the velocity width $W_{20}$, corrected for velocity dispersion, and an estimated inclination angle of $i = 68\degr$, we derive V$_{\rm rot}$ to be 19~\kms.  The radius of \HI\ gas, as seen on the MOM0 maps of Figure~\ref{fig:HImaps1}, is $\sim$ 1.2~kpc.

Due to significant disturbances, the velocity field of this galaxy is not suitable for deriving a rotation curve. Instead, we use the estimated radius and corrected velocity to calculate the indicative dynamical mass. Applying equation \ref{eq:2}, we find $M_{\rm dyn} \sim 1.5 \times 10^{8}$~M$_{\odot}$.
%Here, both $R$ and $V$ are directly derived from observations and corrected for the inclination angle $i$, estimated with formula \ref{eq:1}.  $\cos$(i) is derived via the relation above from the ratio of  minor to major axis $b/a$. We adopt for further the inclination angle of $i$=68\degr\   as presented in Table~\ref{tab:param_GMRT},
% that is multiplicat. factor sin(i)^-2 *cos(i)^-1 = 2.666
% NEED to examine the possible range of i, based on elongation of HI body
% the real V_rot = V_obs/sin(i); real R = R_obs/cos(i)   that is we apply factor 1/sin(i)^2*1/cos(i) = 2.666
% E.g. b/a of 0.5 and assuming an intrinsic axial ratio of 0.25),
% 2.3*10^5*1.3*34^2*1.333*2.0
%the velocity range of $\sim$43~\kms\ and the radius R(\HI) of $\sim$1.3~kpc, 
%(at the \HI\ column density level of1.25~10$^{19}$~\atoms), \
Comparing the \HI\ profiles from ALFALFA and GMRT in Fig.~\ref{fig:HI_prof}, it is evident that the larger beam of GMRT significantly resolves out the low-density outer layers of this galaxy. Consequently, the value of M$_{\rm dyn}$ could be larger.
 
\subsubsection{J0231+2542=AGC122400}
\label{ssec:J0231}

This XMP dwarf also was discovered as an \HI\ source in the blind \HI\ survey
 ALFALFA \citep{Haynes18} and identified with a faint optical galaxy. 
%Its integrated \HI\ parameters will be compared with those derived from the GMRT data in Section~\ref{sec:dis}.
This very LSB dwarf shows elongation in its optical body approximately in the S--N direction (PA $\sim$ 0\degr) (see Figure~\ref{fig:FCharts}). A faint blue `flow' from the North edge to the West suggests some tracers of past or current disturbance. 
%As shown in the top row left panel of Figure \ref{fig:HImaps2}, where the \HI\ contours, at a resolution of $\sim$ 40$\arcsec$ (43.5$\arcsec$ $\times$34$\arcsec$) are overlaid on the optical $g$ band image, the major axis of the \HI\ body is clearly inclined at PA $\sim$ --40\degr. The total extent of \HI\ is, however, comparable to the beam width, so that the disturbed morphology and velocity fields are largely smeared by the beam in these MOM0 and MOM1 moment maps.

The top left panel of Figure \ref{fig:HImaps2} shows the \HI\ contours at a resolution of approximately 40$\arcsec$ (43.5$\arcsec$ $\times$34$\arcsec$) overlaid on the optical $g$ band image. The total extent of \HI\ is comparable to the beam width, causing the disturbed morphology and velocity fields to be largely smeared in the MOM0 and MOM1 moment maps.

In the second  and the third rows of Figure \ref{fig:HImaps2},
we present similar \HI\ flux maps overlaid onto the optical image, with FWHM beam widths of $\sim$15$\arcsec$ (16.5$\arcsec$$\times$13$\arcsec$) and $\sim$11$\arcsec$ (12$\arcsec$ $\times$9.5$\arcsec$), respectively. In these maps, the \HI\ body appears to consist of two approximately equal parts with density peaks on both sides of the stellar body. The major axis of the entire \HI\ body, based on its MOM0 map, is perpendicular to the optical major axis. However, we note that the orientation of the large-scale velocity gradient in the velocity field, i.e., the kinematical major axis of the galaxy, matches the optical orientation.

Therefore, we see two signatures of the non-equilibrium state of gas in this galaxy. The first indicator is that the optical orientation differs from the major axis based on \HI\ morphology, but matches the kinematic major axis.
%The first one is related to the large misalignment of the major axes of the stellar and gas components.
The second signature is the non-equilibrium state of the gas body itself. This is well seen in the maps with the beam widths of $\sim$15$''$ and $\sim$11$''$. The signatures of the latter are both, the odd morphology and the irregular velocity field.  However, despite the clearly seen disturbed state of the \HI\ gas,  the global morphology and the velocity field indicate that at first approximation the velocity field has the dominant direction of its gradient, roughly from the S to the N, with the full range of radial velocities of $\lesssim$20~\kms\ that can be attributed to the total angular momentum.

In the right-hand panels we present the maps of the observed velocity dispersion $\sigma_{\rm V}$ (MOM2).
They reach the maximal values of 8--11~\kms\ near the central part of the \HI\ body, which, in turn, is close to the position of the stellar body.  After correcting the observed values for channel width and velocity gradient (similar to that shown for J0001+3222), we derive the maximal values of $\sigma_{\rm true} \sim$ 9~\kms.

The analysis of its environment up to the distances of $\sim$30$\arcmin$ ($\sim$130~kpc) was performed with HyperLEDA and on the deep colour images of the Legacy survey,  within a square region of 30$\arcmin$ $\times$ 30$\arcmin$, somewhat larger than the primary GMRT beamwidth FWHM diameter ($\sim$22$\arcmin$ or $\sim$100~kpc). This search reveals three bluish irregular galaxies at the angular distances of $\sim$1.3$\arcmin$, 4.5$\arcmin$ and 10$\arcmin$. These potential companions/disturbers, with the total B-magnitudes fainter than that of J0231+2542 by $\sim$2.8, 2.1 and 1.5~mag, are situated at the projected linear distances of 5.7, 20 and 45~kpc, respectively. We search for the \HI\ counterparts of these galaxies in our GMRT observations.
% AGC122400 (37.8415, 25.7123, B=18.92)
% N1 (37.8290, 25.7294, B=21.75, d=1.27'; => 5.7 kpc, dB~2.83 mag);
% N2 (37.7709, 25.6886, B=20.90, d=4.5' = 20.1 kpc; dB~2.08 mag);
% N3 (37.8034, 25.8742, B=20.41, d~10' = 45.1 kpc; dB = 1.5 mag)   SDSS J023112.80+255227.2
% disturbers with mass and luminosity comparable or larger than the studied XMP dwarf. 

If these bluish galaxies have radial velocities within $\pm \sim$380~\kms\ of the velocity of J0231+2542, one could, in principle, detect the brightest and the most distant object, SDSS J023112.80+255227.2, at a low S-to-N ratio, assuming a gas-mass to blue luminosity ratio of M(\HI)/ L$_{\rm B}$ $\sim$ 1. However, nothing hints at a signal at this position in the data cube.  Two other bluish galaxies are too faint for detection.

The use of the available velocity field for the estimate of the dynamical mass in this object is also limited by the complex gas motions and morphology. Following the approach in J0001+3222 from the previous section, we estimate the indicative dynamical mass using Formula~\ref{eq:2}. We use the parameter W$_{\mathrm 20}$ = 31~\kms\ and adopt inclination angle $i$ = 49\degr\ and derive its rotation velocity V$_{\rm rot}$ to be 21~\kms\, and  we obtain a dynamic mass  $M_{\rm dyn}$~$\sim$ 4.2~$\times$~10$^{8}$~M$_{\odot}$.

\subsubsection{J0306+0028=PGC1166738}
\label{ssec:J0306}

This galaxy was initially identified in the SDSS \citep{Adelman-McCarthy07} as  a nearby emission-line dwarf SDSSJ030646.86+002810.2
with the radial velocity of V$_{\rm hel}$ = 711~\kms.
Its \HI\ emission was previously detected by \citet{vandriel16} (cataloged as NIBLES0347) at the close radial velocity
of 699~\kms\ with an integrated flux S(\HI) = 0.55$\pm$0.25~Jy~\kms, which matches well with our GMRT observations (see \S \ref{sec:HI_parameters}).

The optical morphology of this galaxy in the deep image of the Legacy survey (Fig.~\ref{fig:FCharts}) appears as a slightly elongated blue body with several knots. Its major axis is oriented approximately in the S--N direction (optical PA $\sim$ 10\degr). In the left panel of the top row of Figure~\ref{fig:HImaps3}, we show the \HI\ contours at a resolution of  $\sim$40$\arcsec$ (43.5$\arcsec$ $\times$36.5$\arcsec$), overlaid on its optical ($g$-band) image taken from the Legacy surveys database \citep{Dey19} in grey scale. %overlaid on the PanSTARRS1 image in grey scale. 
The \HI\ body is elongated, with its density peak close to the center of the optical image. The velocity field, shown with isovelocity contours and in the colour scale, is barely resolved with this larger beam. We discuss it below for maps with the higher angular resolution.

In the middle and bottom rows of Fig.~\ref{fig:HImaps3}, we present  similar maps with the beam widths of $\sim$15$\arcsec$ (17$\arcsec$ $\times$13.5$\arcsec$)  and $\sim$11$\arcsec$ (12.5$\arcsec$$\times$10$\arcsec$).
Here, the asymmetric morphology of the \HI\ body is well seen in the left panels. There are two distinct components of the main body: the brighter one, with the peak, centered roughly at the position of the optical body, and the fainter component, elongated at PA $\sim$ --45\degr. Additionally, there is a separate faint \HI\ component at $\sim$30$\arcsec$ to the South of the center of the  optical body. This small \HI\ blob seems to be kinematically detached from the gas in the main component since the velocity field  clearly breaks at their boundary.
No optical counterpart of this blob is seen, however, to the limit of the available images in the mentioned above deep surveys.

In the right-hand panels we show the maps of the observed velocity dispersion $\sigma_{\rm V}$ (MOM2).
Its maximal observed values do not exceed of 8~\kms. After the corrections, similar to those in the previous dwarfs, the values of $\sigma_{\rm true}$ do not exceed $\sim$6~\kms.

The examination of the environment of this galaxy within the radius of 30$\arcmin$ ($\sim$100~kpc)
with the HyperLEDA and with the Legacy DR10 colour images, reveals only one potential counterpart, a bluish irregular galaxy at the projected distance of 7.45$\arcmin$ (24~kpc), with B-mag of $\sim$20.4~mag. As adopted above for the case of J0001+3222, for the typical value for such galaxies, of M(\HI)/L$_{\rm B} \sim$1, its expected flux F(\HI) is only $\sim$ 0.05~Jy~\kms, well below the detection limit of our observations.
% XMP 46.6970, 0.4691, B=17.30)
% N1  46.8212, 0.4700 (g=20.08,r=19.83, B=20.38, at 7.45' or 24.3 kpc)

Similar to the two previous XMP dwarfs, the apparent morphology and velocity field are complex. Moreover, for the more extended structure, the velocity field is poorly resolved. As in the previous cases, we derive the indicative dynamical mass with Formula \ref{eq:2} above. We adopt the total extent of \HI\ body as seen in the  MOM0 map with the beam size of $\sim15''$, 
which corresponds to the radius of  R = 1.1 kpc. Adopting the parameter W$_{\mathrm 20}$ = 24~\kms\ and the $i$=67\degr, as shown in Table~\ref{tab:param_GMRT}, we obtain the estimate of the rotation velocity of $\sim$14~\kms. The corresponding value of the dynamical mass is 
 $M_{\rm dyn}$ = 2.1$\times$10$^{8}$ M\sunn.

\begin{figure*}
 \centering
  \includegraphics[width=0.95\linewidth]{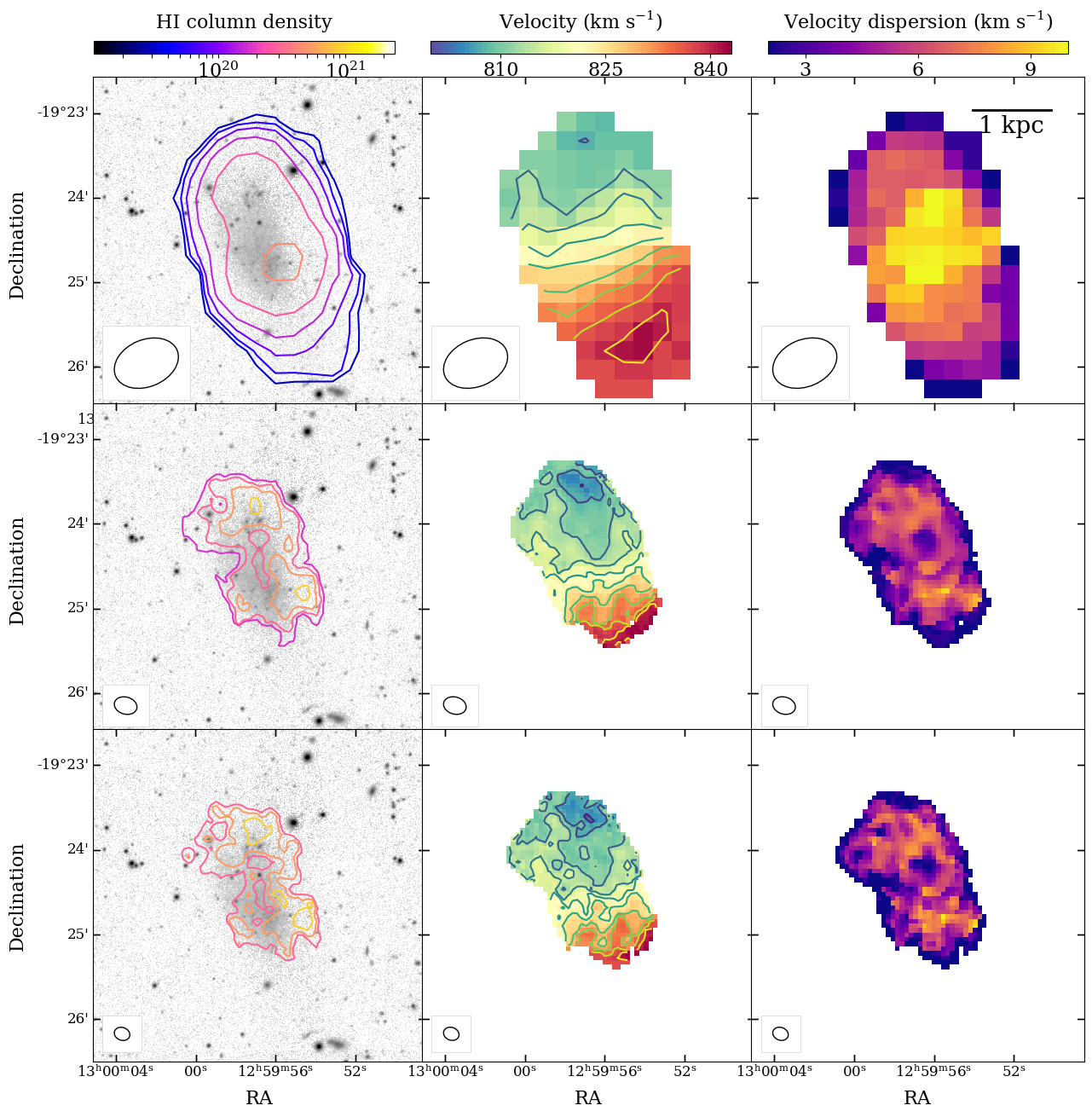}
  \caption{\label{fig:HImaps4} \HI\ moment maps for XMP void dwarf J1259--1924. Panels and symbols as in Figure \ref{fig:HImaps1}. 
  The \HI\ contours overlaid on Legacy g-band image in left panels begin at 3$\sigma$, i.e. 3.0 $\times$10$^{19}$, 1.7 $\times$10$^{20}$, and 2.9 $\times$10$^{20}$~\atoms\ at  40$\arcsec$', 14$\arcsec$', and 10$\arcsec$ resolution respectively, and increase by factor of $\sqrt{3}$. Isovelocity contours in the middle panels are spaced at 4~\kms}
\end{figure*}

\subsubsection{J1259--1924 = PGC044681 = KK176 = HIPASSJ1259--19}
\label{ssec:J1259}

J1259--1924 was first identified as a nearby galaxy and a gas-rich object
by \citet{Huchtmeier00}. Subsequently, it was also observed in \HI\ line by \citet{Pisano11}, who
categorized this galaxy (listed as SGC 1257--1909) as a member of  a so-called `HIPASS' Group, which includes
three other galaxies. Their \HI\ mapping had a relatively poor angular resolution of 177$\arcsec$ $\times$ 40$\arcsec$.
Therefore, we only can use their integrated \HI\ parameters for comparison.
The nearest neighboring galaxy to J1259--1924 is UGCA320 (DDO161 or J130316.8--172523), with a radial velocity V$_{\rm hel}$ = 742~\kms, situated at a projected distance of approximately $\sim$2.2\degr, corresponding to 0.28 Mpc. UGCA320 is situated at a known TRGB-based distance of 6.03 Mpc \citep{Karachentsev17}, while J1259--1924 has a TRGB-based distance of 7.3 Mpc.  The radial velocity of UGCA320 differs by $\Delta$V = --86~\kms\ from that of J1259--1924.  This difference, assuming the Hubble flow, corresponds to the mutual line-of-site distance of $\sim$1.2~Mpc, which is very close to the difference of distances, determined via the TRGB method. This indicates that two galaxies do not belong to the same aggregate as also was noted by \citet{Karachentsev17}.

The main galaxy of this `HIPASS' group, the  subluminous Sc spiral NGC5068 (J131854.8--210220),
with the absolute magnitude of M$_{\rm B}$ = --18.45~mag (HyperLEDA) and with V$_{\rm hel}$ = 671~\kms, is a part of the NVG sample of galaxies in the nearby voids \citep{pustilnik19}.
Its TRGB-based distance, D=5.15~Mpc, was determined by \citet{Karachentsev17}. This value is  $\sim$2.1~Mpc smaller than the distance to J1259--1924. This indicates that  J1259--1924 is not associated to  NGC5068. This conclusion was also made by \citet{Karachentsev17}.  They made a similar conclusion for UGCA320 and its companion UGCA319.

We present the \HI\ column density and velocity distributions of J1259--1924 in Figure~\ref{fig:HImaps4}.  The top row displays maps with the beam width of FWHM$\sim$40$\arcsec$ (46.5 $\times$ 33.5).
At this low resolution, the \HI\ body (left-hand panel), elongated approximately from the S to the N,
has a rather simple morphology of a disc seen at a moderate inclination angle.  
The velocity field (in the middle panel), in general, corroborates this impression. It looks rather regular and resembles the appearance of the global rotation.

The right panel shows the observed velocity dispersion $\sigma_{\rm obs}$ (MOM2) of \HI\ gas. As the colour scale shows, $\sigma_{\rm obs}$ varies along the body from rather typical level of 6--8~\kms\ in the outer parts to the elevated value of 12~\kms\ in the center. 
To derive the true value of velocity dispersion, we apply the same corrections as for the previous galaxies that gives us the range for $\sigma_{\rm true} \sim$6--10~\kms. 
The regions of the enhanced velocity dispersion are often localized close to the peaks of gas density, and hence, can be related to the sites of the recent or current star formation. 

In the middle and bottom rows we show the higher resolution maps with FWHM of $\sim$14$\arcsec$ (16$\arcsec$ $\times$12$\arcsec$) and $\sim$10$\arcsec$(11$\arcsec$ $\times$8.5$\arcsec$). These maps show only the shrinked portion of the \HI\ body visible in the top row due to `resolving-out effect' of the outer parts, particularly those with the lowest column densities. In these moment maps, the extent of \HI\ is reduced relative to that of the whole body shown in the top row. The integrated flux at 15$\arcsec$ resolution drops to 3.8 Jy km/s from 4.3 Jy km/s at 40$\arcsec$ resolution, which is expected when the emission is resolved out. The morphology of this structure is complex, with multiple local density peaks at column density level of (6--8)$\times$10$^{20}$~\atoms. The respective MOM1 maps show the complex structure. While a general velocity gradient is observable from the north to the south, covering a full range of $\sim$35~\kms, there are regions exhibiting substantial velocity disturbances, with extents comparable to the beam size. Some of these disturbances coincide with the positions of local gas density peaks.

The smooth density map and velocity field with a beam size of FWHM $\sim$40$\arcsec$ are well resolved and appears rather regular, albeit also with tracers of the apparent disturbance. This  map can be used, similar to the previous galaxies, to estimate the indicative dynamical  mass within the limits of the detected \HI\ body. We adopt the linear radius of R(\HI) = 2.2~kpc using the MOM0 map with 15$\arcsec$ resolution. With the full range of radial velocity of W$_{\mathrm 20}\sim$50~\kms, and inclination angle of $i$ = 64\degr, the adopted V$_{\rm rot}$ $\sim$28~\kms. Then, the estimate of M$_{\rm dyn}$  is  5.9$\times$10$^{8}$~M\sunn. 
The dynamical mass would be a factor of 1.4 larger than this if it was derived with R(\HI) obtained on the MOM0 map with the beam width of $\sim$40$\arcsec$.

The local environment of J1259--1924 was examined on the DR10 colour image of the Legacy survey.
Within the radius of 30$\arcmin$, the only bluish galaxy which can be potentially in the LV, is
PGC858037 (J125922.48-191358.1), with B$_{\rm tot} \sim$18.0~mag and A$_{\rm B}$ =0.4, situated
at the projected distance of 13.7$\arcmin$ (28~kpc).
If this galaxy would have radial velocity within the observed range for the target object,
for the adopted above typical ratio M(\HI)/L$_{\rm B} \sim$ 1, one could expect its F(\HI) at the level of 0.6~Jy~\kms.
However, due to its off-set position relative to the center of the GMRT primary beam, one expects a flux loss by
a factor of more than two, and hence, the flux of $\lesssim$0.3~Jy~\kms. If it would have the mentioned
above flux of 0.6~Jy~\kms, and the velocity below $\sim$3000~\kms, it would be detected in the map
of \citet{Pisano11}. 
Our data cubes at various  angular resolutions  do not indicate any signal at
this position.

% J1259-1924: RA Dec=194.985339 -19.411942
%RA Dec=194.843653 -19.232862
%D(26)=0.372 arcmin, b/a=0.55, PA=52.0 deg

\begin{figure*}
  \includegraphics[width=0.95\linewidth]{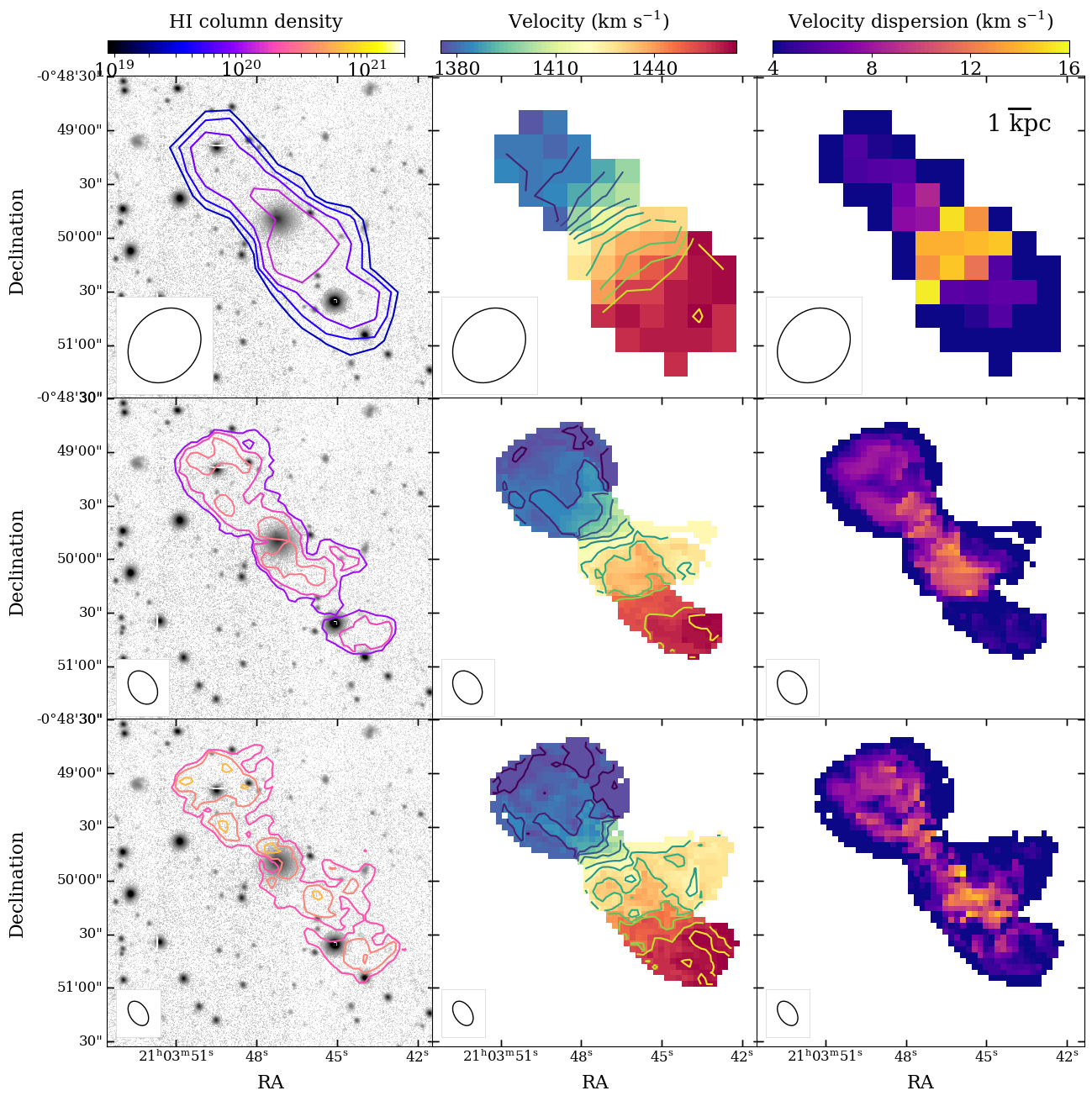}
  \caption{\label{fig:HImaps5}  The \HI\ moment maps for XMP void dwarf J2103--0049. Panels and symbols as in Figure \ref{fig:HImaps1}. The \HI\ contours overlaid on SDSS Stripe82 $g$-band  image in left panels begin at 3$\sigma$, i.e. 2.4 $\times$10$^{19}$, 1.0 $\times$10$^{20}$, and 1.9 $\times$10$^{20}$~\atoms\ at  41$\arcsec$', 17$\arcsec$', and 12$\arcsec$ resolution respectively, and increase by factor of $\sqrt{3}$. Isovelocity contours in the middle panels are spaced at 8~\kms}
\end{figure*}

\subsubsection{J2103--0049=PGC1133627}
\label{ssec:J2103}

The global \HI\ parameters of J2103--0049 were initially obtained by \citet{vandriel16} on
observations with the Nan\c {c}ay  Radio telescope (NRT). The uncertainty  of their \HI\ flux is $\gtrsim$ 20\%.
Therefore,  the better quality data, including the column density maps and velocity field are clearly needed.
It was also quite atypical for that faint and compact dwarf to have rather large velocity
width W$_{\mathrm 50} \sim$ 80~\kms.

\HI\ maps for  galaxy J2103--0049 are shown in Figure~\ref{fig:HImaps5}.
In the top row, the MOM0, MOM1 and MOM2 maps with the beam width of FWHM $\sim$41$\arcsec$
(45.5$\arcsec$ $\times$37.5$\arcsec$) are presented. In the left-hand panel, the \HI\ density contours are
overlaid on the grey scale SDSS Stripe82 optical image in filter $g$. The main \HI\ body is well elongated
(axial ratio of $a/b \sim$3), with the total extent of $\sim$3$\arcmin$ and the major axis directed at PA $\sim$ 41\degr,  as obtained from SoFiA output. This is in contrast with the optical appearance of the almost round blue oval with the total extent of $\lesssim$15$\arcsec$. The  bending \HI\ body is clearly disturbed,  with appendages visible on the W and E sides.
As seen in the middle panel (MOM1), the full range of velocities is $\sim$80~\kms, that corroborates the respective
\HI\ profile width from \citet{vandriel16}. The general velocity gradient is close to the direction along
the `major' axis, but the velocity field looks asymmetric. The map of the visible velocity dispersion (MOM2) in the right-hand panel displays the values substantially larger than for the other dwarfs in this work. In particular, this relates to the regions at the  W edge of the main body, where MOM2 is enhanced up to $\sim$20~\kms. The highest values of MOM2 seem to be related  to regions of the large velocity gradients smeared by the large beam size of the lowest resolution  moment maps.

In the middle and bottom rows, we show the similar maps for beams with FWHM of
$\sim$17$\arcsec$ (20$\arcsec$ $\times$15$\arcsec$) and $\sim$12$\arcsec$ (14.5$\arcsec$ $\times$9.5$\arcsec$), respectively.
With this 2--3-times higher angular resolution, the overall size of \HI\ body remains close to that for the lowest resolution map.
This implies that the outer parts appear to have sufficiently high densities and are not resolved out by the smaller beams.

The higher resolution maps reveal a three-part structure, with the isovelocity lines bending on the both, the NE and SW edges. 
Besides, the central region of the gas body, surrounding the compact stellar image, can be dynamically separated from the two outer parts, since within this region the gas radial velocity shows only small changes.  

In MOM2 maps, the region of the largest velocity dispersion, of $\sim$15--18~\kms\ falls close to the stellar body of the galaxy, in which we witness the recent and current star formation. With the substantial correction of the observed MOM2 for the 
contribution of the velocity gradient ($\sim$20~\kms\ on the scale of the beam width), the related  values of 
$\sigma_{\rm true}$ are reduced to $\sim$11--12~\kms. This enhanced $\sigma_{\rm true}$ can indicate the local agitation of the surrounding gas.

All the global irregularities, both in the gas distribution and in the velocity field, indicate
the non-equilibrium state of the gas body in this galaxy.

On the lowest resolution maps, with the beam width of $\sim$40$\arcsec$ (Fig.~\ref{fig:HImaps5}, top row), both, the morphology and velocity field resemble at the first approximation  a rotating disc, albeit with the apparent disturbances.  We use both, the gas body extent and the full range of the monotonic velocity gradient to get the estimate of the indicative dynamical mass, similar to other studied dwarfs.
The angular extent along the major axis of the \HI\ body is $\sim$3$\arcmin$ (or 14.6~kpc), and the related radius is $\sim$7.3~kpc.
%At the adopted distance to this galaxy, this corresponds to the linear radius of 8.86~kpc. 
The full velocity range on the MOM1 map is $\sim$90~\kms.  Using the parameter W$_{\mathrm 20}$ $\sim$ 98~\kms\ (see Table~\ref{tab:param_GMRT}),  inclination angle $i$~=~84\degr, we derive the  estimate of V$_{\rm rot}$ = 46~\kms. The corresponding dynamical mass is of M$_{\rm dyn}$ = 7.8$\times$10$^{9}$~M\sunn.

We examined the local environment of this galaxy in order to identify potential disturbers.
Within a radius of 30$\arcmin$, HyperLEDA provides the only galaxy with a close radial velocity.
This is PGC1139658 = J2104--0035, at V$_{\rm hel}$ = 1401~\kms\ \citep{Ekta08} and
B$_{\rm tot} \sim$ 17.4~mag, located at a projected distance of 117~kpc ($\sim$23$\arcmin$). 
%  PGC1133627  J210347.2-004950 1411
%  PGC1139658  J210455.3-003522 1395
%  d(RA) = 68s = 17 arcmin   d(Dec) = 14.47 arcmin   NRT beam 3.7' x 22'  background offset 15' to East FHI=2.0
This well-known gas-rich XMP dwarf \citep{Izotov06,Ekta08,Pustilnik20a} is unbound
to the target galaxy J2103--0049, but appears to be a part of the same filament of the void substructure.

In the Legacy survey image, we identify two faint bluish galaxies: one located at $\sim$1.6$\arcmin$ ($\sim$8~kpc) to NW,
with B$_{\rm tot} \sim$20.6~mag, and another at $\sim$2.9$\arcmin$ ($\sim$15~kpc) to S, with B$_{\rm tot} \sim$19.7~mag.
 Adopting as above for the other void dwarfs, the typical ratio of M(\HI)/L$_{\rm B} \sim$1, their expected
F(\HI) are only $\sim$0.04 and 0.11~Jy~\kms, respectively, which are too low for the detection with the current GMRT observations. An alternative option is to obtain their optical redshifts, which appears to be feasible.
If these faint galaxies are situated in the same `cosmic cell' as the target galaxy, their baryonic masses are a factor of $\sim$25 and $\sim$10 lower than that  of  the target object. Accounting for their  projected
distances, these low-mass objects at the 3D distances of $\sim$12 and $\sim$20~kpc, could, in principle have a sizable
tidal effect on to the main, J2103--0049. Thus, if their proximity will be confirmed by the optical spectroscopy, a more detailed analysis of their possible interaction will be required. 
% Thus, one does not expect their sizable disturbing effect on the main galaxy.
% our target: RA,Dec = 315.9468, -0.8303
% Mags1:  g=20.28, r=19.88, i=19.61, z=19.57 ==> B~20.60
% Source1     RA,Dec = 315.9324, -0.8104  J210342.80-004837,    1.1'& 1.2' = 1.6 arcmin
% Source2 Mags: g=21.01, r=20.52, i=20.28, z=20.20  => B~21.4
%             RA,Dec = 315.9393, -0.8178  J210345.42-004904                  0.87 arcmin
% Source3     RA,Dec = 315.9461, -0.8793                                     2.94 arcmin to S
% CModel mags:  g=19.36, r=19.01, i=19.08, z=18.83  => B~19.7 B0~19.4 ==> FHI ~0.114

\section{DISCUSSION}
\label{sec:dis}

\subsection{Global parameters of void XMP dwarfs}
\label{ssec:global_parameters}
In this paper, we present the GMRT \HI\ mapping of the first 5 XMP void dwarfs from a subsample
of 15 isolated such galaxies. A comprehensive statistical analysis of their main properties, along with a comparison to the other known dwarf samples with the similar range of baryonic masses, such as FIGGS \citep{Begum08a} and SHIELD \citep[][and references therein]{McQuinn21}, is  planned after presenting the remaining 10 void dwarfs in this category. Nevertheless, it is worth briefly describing and summarising their main properties in Tables \ref{tab:param_literature} and \ref{tab:param_GMRT}.
At first approximation, they appear to be representative of the larger sample of void XMP dwarfs, mapped so far in the \HI\ 21-cm line.

The parameters for the studied XMP dwarfs in Tables~\ref{tab:param_literature} and \ref{tab:param_GMRT} show a rather small scatter in their global parameters. This can be partly attributed to selection criteria used to separate such objects from the full list of 1350 objects in the NVG sample, as described in \citet{Pustilnik20a}. For example, their absolute blue magnitudes all fall within the range M$_{\rm B}$ of --12.4 to --14.1~mag, that is their luminosity varies by a factor of about five. All these dwarfs are gas-rich, with the range of their  M(\HI)/L$_{\rm B}$ $\sim$ 1--7.

The atomic gas (\HI\ and He) is the dominant component of their baryon mass.  M$_{\rm gas}$ varies by a factor of $\sim$7, ranging 
from 2.5 to 17.4 $\times$ 10$^{7}$~M\sunn. This roughly reflects the range of their optical blue luminosity, provided mostly by a relatively young stellar population, as indicated by their bluish or blue integrated optical colours.
Since the apparent magnitudes of these dwarfs are rather faint, not all of the galaxies have good optical multi-colour
photometry and NIR magnitudes. Therefore, the estimates of their stellar mass are available only for a subset
of these XMP void dwarfs. For the four dwarfs with available estimates of the stellar mass, it comprises roughly from 5 to 10 percent of the total baryon mass.

The width of the line profile of the integrated \HI\ emission serves an indicator of the amplitude of gas motions. The line profile, besides the systematic motions (rotation, expansion, accretion), includes as well the contribution from both, thermal and turbulent gas motions in the galaxy, with the typical velocity dispersion of 6--10~\kms \citep[e.g.][]{Tamburro09}. The profile widths -- W$_{\mathrm 50}$ and W$_{\mathrm 20}$ for our GMRT data, as derived from the lowest angular resolution cubes (FWHM $\sim$40$\arcsec$), are shown in Table \ref{tab:param_GMRT}. The derived values of W$_{\mathrm 50}$  are relatively small, ranging from $\sim$13 to $\sim$40~\kms, except for galaxy J2103--0049, for which this parameter reaches of 86~\kms. This galaxy is the most distant and massive of all five dwarfs in this sample. One of the possible reasons of its increased velocity amplitude and enhanced velocity dispersion, apart from the higher mass, can be a merger, the past disturbance by a fly-by comparable mass dwarf, or the recent gas accretion from the cosmic filaments, as the gas morphology and kinematics hint (see Fig.~\ref{fig:HImaps5}), and as is discussed in more details below.

The  estimates of \HI\ profile widths, presented in Table~\ref{tab:param_GMRT}, allow one to estimate the value of V$_{\rm rot}$. For all dwarfs except J2103-0049, the derived values of V${\rm rot}$ fall within the range of 19 to 31~\kms.
%23 to 47~\kms. (I updated these numbers from the fresh versions in the descriptions of individual dwarfs)
The indicative dynamical masses in Table~\ref{tab:param_GMRT} were estimated using two methods: the radius corresponding to the surface density level of 1M\sunn pc$^{-2}$ and the radius based on the extent of MOM0 map (at 3$\sigma$). These measurements were derived for all galaxies using MOM0 maps with a beam width of $\sim$15$\arcsec$. At this resolution, all studied galaxies are sufficiently well resolved, and the respective radii are reliably measured. It is worth noting that the estimate of the indicative dynamical mass in formula \ref{eq:2} assumes regular rotation. However, since we observe varying degrees of evidence for an `unsettled' state of gas in all our dwarfs, the basic assumption underlying formula \ref{eq:2} may not hold true. Consequently, the derived values of dynamical mass are likely subject to additional uncertainties.

The dynamical mass, defined in this manner, falls within the range of (1.6--9.3)$\times$10$^{8}$~M\sunn. The case of J2103--0049 appears to be an exception. For this object, with V$_{\rm rot}$ = 46~\kms, M$_{\rm dyn} \sim$ 8$\times$10$^{9}$~M\sunn, is about an order of magnitude larger than for the rest of the group. 

 On several parameters, like the total baryonic mass, J2103--0049 appears somewhat extreme relative to the rest of the group, having this mass of about twice larger  than for J1259--1924. When we include in the comparison the dynamical mass,  this galaxy appears  significantly outlying, exceeding J1259--1924 by more than an order of magnitude. The large difference in their dynamical masses is primarily caused by a factor 
of $\sim$3.3 larger radius, and by a factor of $\sim$2 larger the \HI\ profile width, which enters in quadrature.

The latter difference can be related to its probable merger case, since for this interpretation, the observed velocity range could be due to the relative orbital velocity of the two merging components. Another possible scenario for J2103--0049 is the accretion of the metal-poor gas from gaseous filaments. In the case of accretion of the unprocessed gas, one can detect the related drops in the galaxy gas metallicity \citep[see, e.g.,][and references therein]{Sanchez2014, Sanchez2015}. In this context, J2103--0049 reveals the largest (of all five XMP dwarfs in this group) negative deviation of gas oxygen abundance (by $\sim0.6$~dex) with respect of the reference ``metallicity -- luminosity'' relation \citep{Berg12}, in line with the accretion scenario. Moreover, despite the very elongated \HI\ disk, in optics the galaxy has an unperturbed appearance, which also can be explained in the case of the cold gas accretion. Besides, it would explain the dynamical detachment of the central region of the gas body from the two outer parts (see Sec.~\ref{ssec:J2103}). A deeper study of this object is needed to elucidate its outlying parameters.

As for the estimates of the indicative dynamical mass in the four other dwarfs, of $\lesssim$10$^{9}$~M\sunn,  these values are rather lower limits due to the usage of the estimated linear radius in Formula $\ref{eq:2}$. In all these objects, this radius
corresponds to the column gas density of $\sim$1.25 $\times$~10$^{20}$~\atoms\ (or to the mass surface density of 1~M\sunn\ pc$^{-2}$). 
In addition, we also present estimates of M$_{\rm dyn}$, in Table~\ref{tab:param_GMRT}, calculated using the maximal radii of \HI\ body observed in MOM0 maps with the beam width of $\sim$15$\arcsec$.

\subsection{Spin misalignment and lopsidedness of \HI\ gas}

The mutual orientation of gas and stellar spins is an important indicator of the
non-equilibrium state of a galaxy, related to the recent interactions, mergers or gas accretion.
In Figure~\ref{fig:FCharts}, we show the optical images of the studied galaxies,
extracted from the Legacy database, which give the impression of their optical morphology and
orientation. We highlight the most reliable cases where a clear misalignment or, conversely, a good 
alignment of the gas and stellar spins is observed.

For galaxies J0001+3222, J0231+2542 and J1259--1924, their optical low surface brightness (LSB) bodies are
significantly elongated. Their \HI\ gas bodies are clearly elongated as well.

In the case of J0001+3222, the major axis of the \HI\ distribution for the higher resolution maps is elongated approximately along the optical major axis. However, in addition, there is an appendage in the direction,
approximately perpendicular to the optical major axis. %One can suggest that this feature represents a residual of the mainly resolved out extended gas which is misaligned with the optical `disc'.

For J0231+2542, the direction of the major axis of \HI\ distribution is
roughly perpendicular to that of the major axis of the LSB optical body. However, the kinematic major axis of the \HI\ component is approximately aligned with the major axis of the optical body.

The stellar LSB body of J0306+0028 looks roundish and rather irregular due to several brightness spikes.
This precludes the determination of the major axis orientation. Therefore, despite the \HI\
body is well elongated at the lowest resolution, and its orientation keeps approximately the same
for maps with the higher resolution, it is difficult to judge about the relative orientation of spins
for the stellar and gas bodies. Thus, the indication  on the non-equilibrium state of this galaxy
comes primarily from the complex morphology and velocity field of \HI\ gas, which are better seen
in the higher resolution maps in Figures~\ref{fig:HImaps3}.

For the nearest dwarf J1259--1924, both the stellar and the atomic gas bodies are elongated approximately in the same direction.
The \HI\ disc has the complex structure in the MOM0 map with the lowest resolution, with the appendages at the SW edge and on the eastern side. This makes difficult  to determine the alignment between the spins of the stellar and gas discs. At the higher angular resolution, with even more complex morphology, one does not notice an apparent misalignment of  \HI\ body with that of the stellar component. Nevertheless, one can notice one more evidence of the unsettled gas disc appearing as the lopsidedness of the velocity field (MOM1 maps), that is a clear bending of the velocity gradient from the S to N edge. This is best seen in the 40~arcsec beam map, but also is traced in the higher resolution maps.

The void XMP dwarf J2103--0049 also has the roundish form of the optical LSB outer body.
However, a small elongation allows one to estimate the PA of the major axis of $\sim$60\degr.
This seems to differ by $\sim$20\degr\ from the orientation of the \HI\ major axis. As discussed above, the unusual gas morphology and velocity field could indicate an advanced merger. However, the option of cold gas accretion can be even more attractive. To reach more definitive conclusions about the real scenario resulting in the observed disturbed morphology and gas motions, a more detailed analysis of the studied dwarfs is required. This includes multiwavelength data and clear criteria to distinguish between a merger and the secular/intermittent accretion of the unprocessed void gas.

 In fact, the misalignment of stellar and gaseous spins (both, atomic and ionised), including their
counter-rotation, is not a very rare phenomenon. E.g., a large review of the known cases of the cold gas accretion to spiral galaxies was presented by \citet{Sancisi08}. However, the majority of the known cases are related to rather massive discs, where the ordered motions of stars and gas are clearly traced. They include, in particular,  the so-called polar-ring galaxies \citep[e.g.][]{Moiseev_2012,Moiseev_2015,Egorov_2019} and many cases of S0 galaxies \citep[e.g.][]{Silchenko_2019,Silchenko23}.

A fraction of misaligned spins found in the systematical studies with the `SAMI' and `MaNGA'
instruments,  covering  large representative samples of disc galaxies \citep[e.g.][]{SAMI2022,Xu22,Zinchenko23}, 
appears at the level of $\sim$5--12\%. 
The most straightforward interpretation of this phenomenon is an appearance of the recent gas accretion or a minor merger. This occurs when the spin of the main component is inclined relative to the direction of the orbital angular momentum of the inflowing  gas or of the infalling `companion'.

For dwarfs, this kind of phenomenon is noticed and described only for a few objects  \citep[e.g. MRK996,][]{Jaiswal20}. 
For dwarfs in the void environment, this phenomenon is even less studied, mainly due to the absence of suitable void galaxy samples, in which the atomic gas component can be mapped with the sufficient angular resolution and sensitivity. 

For void dwarfs, these should be in principle the same mechanisms, however with the substantially scaled down parameters. In particular, due to the reduced number density  n$_{\rm gal}$, roughly by a factor of $\sim$5 relative to the mean density, typical of walls, the galaxy collision rate can be reduced by a factor of n$^{-2}_{\rm gal}$. In reality, due to the small-scale clustering, this effect should be smaller, however. 
Similarly, as argued by  \citet{aragon13}, the gas motions in the small void filaments are expected to be sub-sound in difference to those for the denser and longer filaments. This can change the parameters of cold gas accretion. 

The known example of accretion and interaction in voids is the polar-ring galaxy VGS\_12 from the Void Galaxy Survey \citep{kreckel12}, 
which resides in the wall between the two prominent voids \citep{Stanionik09}. Similar to the case of J2103--0049, it reveals well elongated \HI\ body and strong misalignment between the gaseous and optical bodies. The authors interpret this misalignment, together with the galaxy's location and its orientation (with the stellar disk oriented roughly parallel to the surrounding wall and the \HI\ disk perpendicular to it), as an indication of gas accretion from out of the voids.
Its gaseous oxygen abundance is also $\sim$0.7~dex lower than expected for its luminosity (Egorova et al., in preparation), which strongly supports the previously suggested scenario of pristine gas accretion.
%and a void triplet VGS-36a,b,c \citep{Beugy2014}. 
Several void XMP galaxies, studied earlier with GMRT, show the clear traces of merging or probable accretion \citep{Ekta08,chengalur17}.

\begin{figure*}
  \centering
\includegraphics[width=0.32\linewidth]{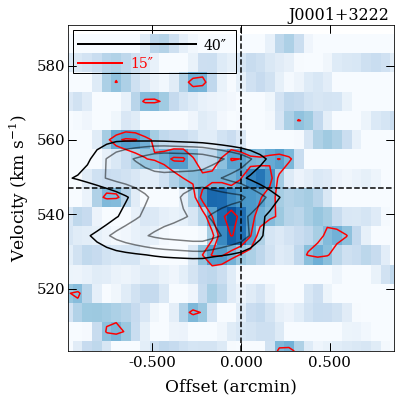}
\includegraphics[width=0.32\linewidth]{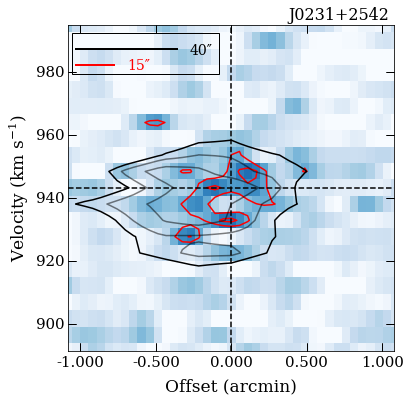}
\includegraphics[width=0.32\linewidth]{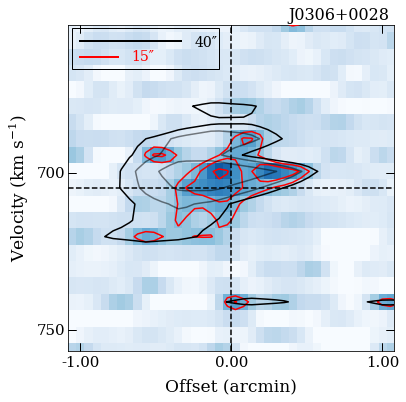}
\includegraphics[width=0.32\linewidth]{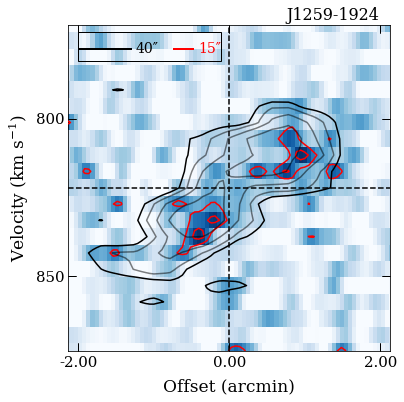}
\includegraphics[width=0.32\linewidth]{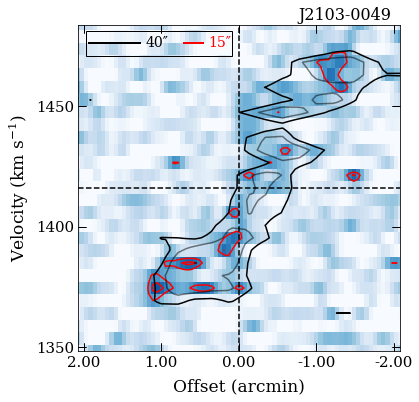}
  \caption{Position-velocity (PV) diagrams at the resolution of 15$\arcsec$ along the major axis. Contours begin at 2.5$\sigma$ and increase by factors
of 1.5. The black contours represent the PV diagrams at 40$\arcsec$ resolution, while the red contours correspond to 15$\arcsec$ resolution. The dashed lines represent the systematic velocity and optical centre of the galaxy. The legend includes 15$\arcsec$ and 40$\arcsec$ bars that show the physical sizes corresponding to these resolutions on the PV diagram. }
  \label{fig:PV}
 \end{figure*}

\subsection{PV Diagrams}
 In addition to the moment1 maps illustrating the velocity fields of the isolated void XMP galaxy sample, we present the position-velocity (PV) diagrams for all five dwarfs in Fig.~\ref{fig:PV}. The black contours represent the PV diagrams at 40$\arcsec$ resolution, while the red contours correspond to 15$\arcsec$ resolution. These diagrams allow us to closely examine the organized motions within these dwarfs and to compare them with the similar published PV diagrams for gas-rich dwarfs, such as those in \citet{Lelli14}. In their study, the diagrams are categorized into `regularly rotating,' `kinematically disturbed,' and `unsettled' groups.

Despite the poor signal-to-noise ratio and the limited spatial resolution for galaxies J0001+3222, J0231+2542, and J0306+0020, their PV diagrams do not show evidence of regular rotation. Instead, they exhibit patterns indicative of the unsettled velocity fields (J0001+3222, J0231+2542, J0306+0020) or kinematic disturbances (J1259--1924).

For J0001+3222, the PV diagram with the 15$\arcsec$ resolution suggests the presence of two systems with very different extents moving in the opposite directions. The PV diagram for J0231+2542, also with a 15$\arcsec$ resolution, appears to trace a shell-like structure.
The galaxy J1259--1924 shows the regular rotation, however it reveals the asymmetrical kinematics in the outer regions.  Additionally, the gas density shows a local minimum at the center, with two dense regions in the middle parts on either side of the disc.

For galaxy J2103--0049, the PV diagram indicates `unsettled' motions, characterized by a steep velocity gradient near the center, which is displaced relative to the roundish optical body, and the constant radial velocity in the outer parts. The unusual feature of these almost constant velocity flows is the presence of dense clumps near their edges, which is atypical for discs with radially decreasing density.

\subsection{Disturbed \HI-morphology in void XMP dwarfs in the context}

As mentioned above, all XMP void dwarfs in this study exhibit
clear disturbances in both, morphology and velocity fields across various spatial scales.
However, no sufficiently close companions were known for any of these five
XMP dwarfs. Therefore, one of the secondary objectives of our GMRT \HI-mapping was to search for
possible dwarf disturbers in the surroundings of the target galaxies.
Our analysis of potential counterparts for the target  dwarfs within the GMRT primary
beam field of view (FWHM $\sim$ 22$\arcmin$) did not reveal any candidates
with \HI\ flux larger than 10--20 percent of that for the main target object. Therefore,
the fly-by tidal effects from the unknown sufficiently close neighbours appear, in general,
unimportant for the studied dwarfs.

These void XMP dwarfs represent the lower parts of the galaxy luminosity or mass functions.
Several samples of low mass dwarfs were studied so far via \HI\ mapping, including SHIELD \citep{McQuinn21}
and FIGGS \citep{Begum08b, Patra2016}. They closely match the mass and luminosity range of the galaxies in this work.
Moreover, by chance, some of the representatives of these samples fall into nearby voids and belong to
our primary Nearby Void Galaxy (NVG) sample \citep{pustilnik19}.  We are planning to compare their properties with our low-mass void sample in upcoming papers, where we will present GMRT observations for the remaining 10 XMP dwarfs. This approach will enable us to obtain statistically significant results. 

 However, it is worth making brief comments on the incidence of the disturbed state of gas in other known samples of late-type dwarfs. In particular, in the sample of 73 late-type dwarf galaxies studied within the framework of the WHISP project \citep{Swaters2002}, evidence of lopsidedness was found in about half of the sample. However, the authors indicate that in 5 cases, this is clearly related to interaction. Additionally, they note that many of their dwarfs are members of small groups and, hence, could likely have undergone interactions in the past.

The sample of 76 faint dwarfs from the FIGGS project \citep{Begum08b, Patra2016}, mostly residing within the Local Volume, is rather similar in its global parameters to our sample in this work. Of them, roughly 40\% reside in groups, $\sim$40\% in the general field, and $\sim$20\% in voids \citep[belonging to the NVG sample from][]{pustilnik19}. In groups, according to our visual inspection, $\sim$80\% of dwarfs with available \HI\ maps show disturbed morphology and kinematics. In the general field (not in voids, but unbound to groups), $\sim$55\% of dwarfs with available \HI\ maps show disturbed morphology and \HI\ kinematics. In voids, of 16 isolated dwarfs with available \HI\ maps, at least 13 ($\sim$80\%) have disturbed \HI\ morphology.

\cite{kreckel12} found that many of the $\sim$60 galaxies in their Void Galaxy Survey sample have strongly disturbed \HI\ gas morphology and kinematics. The authors interpret this as signs of ongoing interactions and gas accretion. Despite their void galaxy sample being significantly biased towards higher mass and higher metallicity objects compared to the XMP dwarfs presented in this paper, their findings suggest that many representatives of the void galaxy population are still in the process of assembling.

 There are several scenarios that could be responsible for the observed non-equilibrium state of the atomic gas in the discussed dwarfs. They include: 1) the effect of localized star formation leading to the formation of expanding gas shells; 2) the result of an advanced stage of a (minor) merger. See the deep optical imaging of the nearby dwarfs in \citet{2024arXiv240601683S} for the firm evidence of their incidence; 3) tidal disturbance by a nearby `dark' neighbour,
that is by a very gas-rich Dark Matter (DM) halo with a small fraction of stellar mass, which could remain undetected in the  moderate depth optical imaging surveys. 4). One more scenario is a long-ago close fly-by passage of a comparable mass galaxy, which then had enough time to travel to the distances of more than $\sim$100~kpc. This takes model simulations in order to better constrain the possible range of impact parameters and the characteristic time during which this type disturbance will leave detectable. And finally,  5) the on-going accretion of intergalactic gas, presumably from the cosmic filaments.
%can act as a disturbance factor.

For shells or outflows related to the strong star formation (SF) burst and their probable cumulative effect, 
one does not expect counter-rotating gas or other global effects. We can not exclude effect of the
old sufficiently powerful localized SF  on the overall gas morphology and kinematics.  However, since the 
majority of our  void dwarfs belong to the LSB population, the alternative scenarios seem to be more probable.

Of them, the tidal disturbance by a fly-by `dark' object is an open option, since the estimate of its effect depends on the observational upper limit on its \HI\ mass and the related mass of the DM halo, on the one hand, and the distance of such `dark' object from the target dwarf, from the other hand. This requires high resolution numerical simulations of small galaxy interactions, which also should account for  baryonic component of the main galaxy being almost purely gaseous. Similarly, the long-lived disturbance in a void isolated dwarf, which occurred after a long-ago fly-by tidal interaction with a similar mass dwarf (which currently looks to be too distant), remains an open question, since it takes a fine adjusting of the observed parameters of the both counterparts.

The remaining two scenarios, of a (minor) merger and of the cold accretion of intergalactic gas, are both quite probable.
However, due to the significantly reduced average galaxy density in voids, the rate of mergers and other variants of collisions is expected to be many times lower than in the adjacent walls and the denser elements of structure. At the same time, as shown by \citet{aragon13}, the specific conditions of the subsonic gas flows in small void filaments, are conducive for the cold gas accretion onto void galaxies. Therefore, one can expect, that if the phenomenon of the ubiquitous non-settled state of gas in the {\it isolated} void dwarfs is real, the main mechanism of the external disturbance of such a galaxy is related to cold gas accretion.

To distinguish between the two scenarios and to understand which one is applicable in specific instances, 
a detailed investigation through advanced simulations is required. The parameters of the studied void 
XMP dwarfs emphasize the necessity for high mass resolution (preferably better than $\sim$10$^5$~M\sunn\ 
on baryonic matter), particularly for these predominantly gaseous objects.

\section{Summary and conclusions}
\label{sec:summary}
We present the results of the sensitive and high-angular-resolution GMRT \HI-mapping of the five most metal-poor (12+$\log$(O/H) = 7.13--7.28~dex) {\it isolated} gas-rich void dwarf galaxies. This work presents the first part of the whole sample of 15 XMP void dwarfs  mapped to date in the \HI\ line with GMRT.
The statistical analysis of these results and the comparison with samples of similar mass and luminosity dwarfs studied by other groups will be addressed in future work. Summarising the results and discussion above, we draw the following
conclusions:
\begin{enumerate}
\item
Despite their isolation from potentially disturbing galaxies, all of them show overall disturbed \HI-morphology and velocity fields, clearly indicating a non-equilibrium state of their dominant baryonic component. Additional examination of the surroundings of the studied dwarfs, conducted on the Legacy survey
images up to projected distances of $\sim$100~kpc, did not reveal potential disturbers. Furthermore, our GMRT observations, with a limiting \HI\ column density sensitivity of (1--2)$\times$10$^{19}$~\atoms, did not detect any potential gas-rich disturbers within the primary beam of GMRT, which corresponds to projected distances of $\sim$30--80~kpc.
\item
Another evidence of the  non-equilibrium state of gas in the studied objects is the
significant misalignment between the spins of \HI\  and stellar `discs', as well as between the 
gas body major axis and the direction of velocity gradient and  gas 'disc' lopsidedness.
The simplest interpretation of this phenomenon is related to the dynamical youth, that is the
unrelaxed state after the recent non-coplanar accretion of the intergalactic medium gas blob
or a tidally-induced significant disturbance by an external dwarf.   %gas-rich merger.
\item
The new results add a significant number of isolated void XMP objects in a non-equilibrium state to  a few similar objects known from earlier studies. We discuss possible scenarios responsible for the disturbed \HI\ morphology and kinematics of the studied dwarfs:  
expelling \HI-gas due to the previous substantial localized SF episode, the recent minor merger, the tidal effect
of the nearby `dark' galaxies, or of the past fly-by collision, or the 'elusive' cold gas accretion. While the 
last scenario seems to be better consistent with the available data, new high-resolution simulations of cold gas accretion 
in voids are necessary to make more certain conclusions.
\end{enumerate}

\section*{Acknowledgements}

This paper is based in part on observations taken with the GMRT. We
thank the staff of the GMRT who made these observations possible.
The GMRT is run by the National Centre for Radio Astrophysics of
the Tata Institute of Fundamental Research. The reported study was funded by Russian Science Foundation according to the research
project  22-22-00654. The research of SK was supported by the South African Research Chairs Initiative (SARChI) 
of the Department of Science and Technology and National Research Foundation. 
%The authors acknowledge the support of this work through the
%RFBR-DST (India) grant No.~18-52-45008--IND-a.

 We thank the anonymous referee for their valuable comments and suggestions, which have helped us improve the presentation and content of the paper. Our thanks are to M.P.~Haynes, who provided us with the ALFALFA \HI\ profiles
for part of the galaxies studied in this work.
The authors acknowledge the spectral and photometric data and the related
information available in the SDSS and the Legacy survey databases.
The Sloan Digital Sky Survey (SDSS) is a joint project of the University of Chicago, Fermilab, the Institute for Advanced Study, the Japan Participation Group, the Johns Hopkins University, the Max-Planck-Institute for Astronomy
(MPIA), the Max-Planck-Institute for Astrophysics (MPA), New Mexico State
University, Princeton University, the United States Naval Observatory, and
the University of Washington.
This research has made use of the NASA/IPAC Extragalactic
Database (NED), which is operated by the Jet Propulsion Laboratory,
California Institute of Technology, under contract with the National
Aeronautics and Space Administration.

The DESI Legacy Imaging Surveys consist of three individual and complementary projects:
the Dark Energy Camera Legacy Survey (DECaLS), the Beijing-Arizona Sky Survey (BASS), and
the Mayall z-band Legacy Survey (MzLS). DECaLS, BASS and MzLS together include data obtained,
respectively, at the Blanco telescope, Cerro Tololo Inter-American Observatory, NSF's NOIRLab;
the Bok telescope, Steward Observatory, University of Arizona; and the Mayall telescope, Kitt
Peak National Observatory, NOIRLab.
NOIRLab is operated by the Association of Universities for Research in Astronomy (AURA) under
a cooperative agreement with the National Science Foundation. Pipeline processing and analyses
 of the data were supported by NOIRLab and the Lawrence Berkeley National Laboratory (LBNL).
Legacy Surveys also uses data products from the Near-Earth Object Wide-field Infrared Survey
Explorer (NEOWISE), a project of the Jet Propulsion Laboratory/California Institute of Technology,
funded by the National Aeronautics and Space Administration. Legacy Surveys was supported by:
the Director, Office of Science, Office of High Energy Physics of the U.S. Department of Energy;
the National Energy Research Scientific Computing Center, a DOE Office of Science User Facility;
the U.S. National Science Foundation, Division of Astronomical Sciences; the National Astronomical
Observatories of China, the Chinese Academy of Sciences and the Chinese National Natural Science
Foundation. LBNL is managed by the Regents of the University of California under contract to the
U.S. Department of Energy.

 We acknowledge the use of the ilifu cloud computing facility—www.ilifu.ac.za, a partnership between the University of Cape Town, the University of the Western Cape, the University of Stellenbosch, Sol Plaatje University, the Cape Peninsula University of Technology, and the South African Radio Astronomy Observatory. The Ilifu facility is supported by contributions from the Inter-University Institute for Data Intensive Astronomy (IDIA a partnership between the University of Cape Town, the University of Pretoria, the University of the Western Cape and the South African Radio Astronomy Observatory), the Computational Biology division at UCT and the Data Intensive Research Initiative of South Africa (DIRISA).

\section*{Data Availability}
The raw data underlying this article are available at the GMRT archive.
The data cubes and the moment maps are available on request from the authors (SK).



\bibliographystyle{mnras}
\bibliography{XMP} % if your bibtex file is called example.bib

\label{lastpage}

\end{document}